\documentclass[final,3p,times,twocolumn]{elsarticle}

\usepackage{amssymb}
\usepackage[utf8]{inputenc}
\usepackage{amsmath}
\usepackage{graphicx}
\usepackage{siunitx}
\usepackage{physics}
\usepackage[export]{adjustbox}
\usepackage{subfig}
\usepackage{framed}
\usepackage{color,soul}
\usepackage[flushleft]{threeparttable}
\usepackage{caption}
\usepackage{float}
\usepackage{multirow}
\usepackage{booktabs}
\usepackage{tabularx}
\DeclareUnicodeCharacter{2212}{-}
\DeclareUnicodeCharacter{0308}{-}
\DeclareUnicodeCharacter{0301}{-}
\DeclareUnicodeCharacter{0327}{-}
\DeclareUnicodeCharacter{2061}{-}

\journal{Elsvier}

\begin{document}

\begin{frontmatter}

\title{Quantum Spin Hall Effect and Su-Schrieffer-Heeger Model Implementation in Novel C$_3$N-based Dumbbell Morphologies}
\author[1]{Deep Mondal\corref{cor1}}
\author[2]{Arka Bandyopadhyay\corref{cor1}}
\author[3]{Atanu Nandy}
\author[1]{Debnarayan Jana\corref{ac}}
\ead{djphy@caluniv.ac.in}
\cortext[cor1]{These authors contributed equally to this work}
\cortext[ac]{Corresponding author}
\address[1]{Department of Physics, University of Calcutta, 92 A. P. C. Road, Kolkata-700009, India}
\address[2]{Solid State and Structural Chemistry Unit, Indian Institute of Science, Bangalore 560012, India}
\address[3]{Department of Physics, Acharya Prafulla Chandra College, New Barrackpore, Kolkata 700131, India}

\vskip 0.25cm
\begin{abstract}
Two-dimensional carbon nitride materials have been the center of attention for their diverse usage in energy harvesting, environmental remediation and nanoelectronic applications. A broad range of utilities with decent synthetic plausibility have made this family a sweet spot to dive into, whereas the underlying analytical aspects are yet to have prominence. Recently, using the machinaries of first principles, we reported a family of six different structures C$_3$NX [J. Phys. Chem. C 127 (2023) 18001] with a unique dumbbell-shaped morphology, functionalizing the recently synthesized monolayer of C$_3$N [Adv. Mat. 29 (2017) 1605625]. Here we have critically explored the non-trivial topological phases of the semimetallic Dumbbell C$_3$NX sheets and nanoribbons. Spin-orbit coupling induced gap across the Fermi level, its subsequent tuning via an external electric field, portrayal of band inversion from the Berry curvature distribution and the evaluation of $\mathbb{Z}_2$ topological index using the Wannier charge center (WCC) firmly establishes the traces of topological footprint. The real space decimation scheme and Green's function technique evaluate the underlying spectral information with corresponding transport characteristics. Fascinating features of these quasi-1D systems are observed utilizing the Su-Schrieffer-Heeger (SSH) model where different twisted phases reveal distinct topological signatures even in a low atomic mass system like DB C$_4$N.
\end{abstract}
\end{frontmatter}

\section{Introduction}
\label{intro}
Two-dimensional (2D) group-IV elemental monolayers and their various allotropes have drawn an enormous amount of attention since the realization of graphene in 2004 \cite{geim2007rise}. Over the years, systems such as silicene, germanene and stanene have all been experimentally synthesized \cite{feng2012evidence,chiappe2014two,zhang2016structural,zhu2015epitaxial,ghosal2023review} with the presence of the Dirac cone having large Fermi velocities. Taking into account the spin-orbit coupling (SOC) interaction, all these structures display a finite gap opening at the high-symmetry `K' point about the Fermi level, which in turn provides the possibility for quantum spin Hall (QSH) effect \cite{qi2010quantum,kane2005z,ezawa2014topological} at ambient temperatures \cite{liu2011low}. This interaction strength is usually very weak for materials especially which are composed of light-mass elements. Since then, a pool of efforts has been made to circumvent and realize this zero bandgap nature (without SOC) of graphene with the unobservably small SOC-induced gap \cite{zhou2007substrate,nath2014ab,takahashi2014band,yao2007spin,rybkin2018magneto}. One very worthy technique is to substitute the parts of carbon atoms with nitrogens, resulting in distinct C$_x$N$_y$ structures. Comparable atomic radii of N atoms allow themselves to fit in different positions within graphene which for different compositions and configurations exhibit a wide range of properties with various novel applications \cite{jana2009first,cao2014g,li2015efficient,yang2016tunable,liu2015electronic,datta2020exploring,debnath2021carbon,bao2021carrier,mortazavi2022combined,fawaz2023emerging}. However, topological insulator (TI) states were never been on the scene for carbon nitrides until the arrival of monolayer s-triazine, chemically known as g-C$_6$N$_6$ which displayed stronger SOC than that of graphene and silicene \cite{wang2014topological}. These special quantum states of matter have drawn enormous research interest in condensed matter physics due to their unconventional electronic properties with promising technological applications \cite{hasan2010colloquium}. This topologically protected behaviour of the electronic wavefunction originates from the symmetries inherent in physical systems encapsulated by topological invariants \cite{bernevig2013topological}. In particular, distinct symmetries, namely, chiral, time-reversal, and charge-conjugation symmetry, impart a robust topological response, resulting in the emergence of topologically protected zero-energy edge states. The simplest model to examine the topological behavior is the famous Su–Schrieffer–Heeger (SSH) model \cite{su1979solitons}, describing a one-dimensional (1D) dimerized chain of atoms that can be either in a metallic phase or insulating phase depending on the ratio of the hopping parameters. Despite numerous promising outcomes \cite{hasan2010colloquium}, the recognition and examination of topological insulators have been confined to a handful of known crystal structures \cite{bradlyn2017topological}.\\
Probing these quantum states of various 2D semi-metallic systems whether planar or buckled with different symmetries - hexagonal, rectangular or square by various schemes has spilled out some fascinating physics over the years \cite{jiao2020tailoring,bandyopadhyay2020topology,bandyopadhyay20218,mukherjee2023application}. In 2017, Yang. et al. showed a controllable large-scale synthesis of a 2D crystalline, hole-free carbon-nitride system called C$_3$N, famously called the 2D polyaniline (PANI) through polymerization which carries some excellent features like ultrahigh mechanical strength, moderate indirect bandgap, optical tunability over the entire visible spectrum and low-temperature ferromagnetism \cite{yang2017c3n}. Following these impressive outcomes, a large section of current research is devoted to the further enrichment of this monolayer graphene-like C$_3$N through the tuning of its electronic structure via doping, strain engineering or applying external electric field \cite{he2018boron,nong2020c3n,bafekry2020electro}. Apart from the aforementioned gr-IV buckled hexagonal honeycomb systems, another very different and morphologically distinct model known as the dumbbell (DB) structure has also been proposed \cite{özçelik2014new,ozcelik2015adsorption}. As opposed to the electronic structure of stanene, its DB form showed an inverted band gap at the center of the Brillouin zone and is reported to maintain a nontrivial topology \cite{tang2014stable}. Also, the recently synthesized DB silicene \cite{leoni2021demonstration}, found to be the true ground state of silicene, behaves as a room-temperature topological insulator under anisotropic strain with a gap achievable up to 36 meV \cite{zhang2016dumbbell}. In 2015, Ozcelik et al. reported a shortcoming for this kind of formation that DB structures cannot be achieved if the adatom and the adsorbent both involve a layer of honeycomb carbon atoms \cite{ozcelik2015adsorption}. But later, two different dumbbell forms are obtained from this single layer C$_3$N through the adsorption of carbon (C) atoms with semi-metallic electronic structure \cite{li2017carbon}. In our recent work, we have discovered a whole new family of six distinct DB systems, spontaneously built out of adsorbing some special adatoms on this newly synthesized monolayer of C$_3$N \cite{jana2023spontaneous}. Based on the type and nature of the adsorbates, these systems electronically develop both semi-metallic and semiconducting features which is quite intriguing for both analytic and application purposes.\\
Shedding light on the underlying analytic regime, here we have shown the distinct topological phases of the DB-C$_3$NX sheets through the interplay between the SOC and band gap modulation with an external electric field. Following that, we have further shown the spectral information with the corresponding transmission probability for the quasi-1D ribbon using real space decimation and Green's function formalism. This leads us towards a complete and rigid understanding of the topological signature of these concerned systems. In particular, using the potential SSH model we have shown how two different twisted phases of the nanoribbons show topological inequivalence.
\section{Methodology}
Our first-principles calculations for the Dumbbell C$_3$NX systems were conducted within the density functional theory (DFT) framework using the {\sc quantum espresso} code~\cite{QE-2017,QE-2009}. The kinetic energy cut-offs of $44$ Ry and $46$ Ry were employed for C$_3$NX sheet and nanoribbons, respectively. Besides, ultrasoft pseudopotentials~\cite{PhysRevB.41.7892} are utilized to represent the core electrons in both cases. In our electronic structure calculations, the incorporation of the spin-orbit coupling effect was achieved through the utilization of full relativistic pseudopotentials. Furthermore, to minimize interactions between periodic images, a vacuum region of approximately 20 \AA~was incorporated along the non-periodic directions for both the sheet and nanoribbons. The van der Waals
interactions are included in the stacked structure by using DFT-D3
method. The plane-wave energy cutoff is set to be 500 eV and the energy
convergence criterion in the self-consistency process is 10$^{-6}$ eV. The Brillouin zone was sampled using a uniform $\Gamma$-centered $k$-mesh of $16\times16\times1$ for the sheet and $20\times1\times1$ for the nanoribbons. The phonon dispersion calculations are performed via the finite displacement method using the open-source Phonopy code by considering a $2\times2\times1$ supercell. The thermal stability of these structures
are also examined under ab initio molecular dynamics (AIMD)
in the canonical ensemble (NVT). The temperature of these
systems are regulated by the Nosé−Hoover thermostat.
\section{Structural detail and electronic nature}
\begin{figure*}[ht]
    \centering
    \includegraphics[scale=0.5]{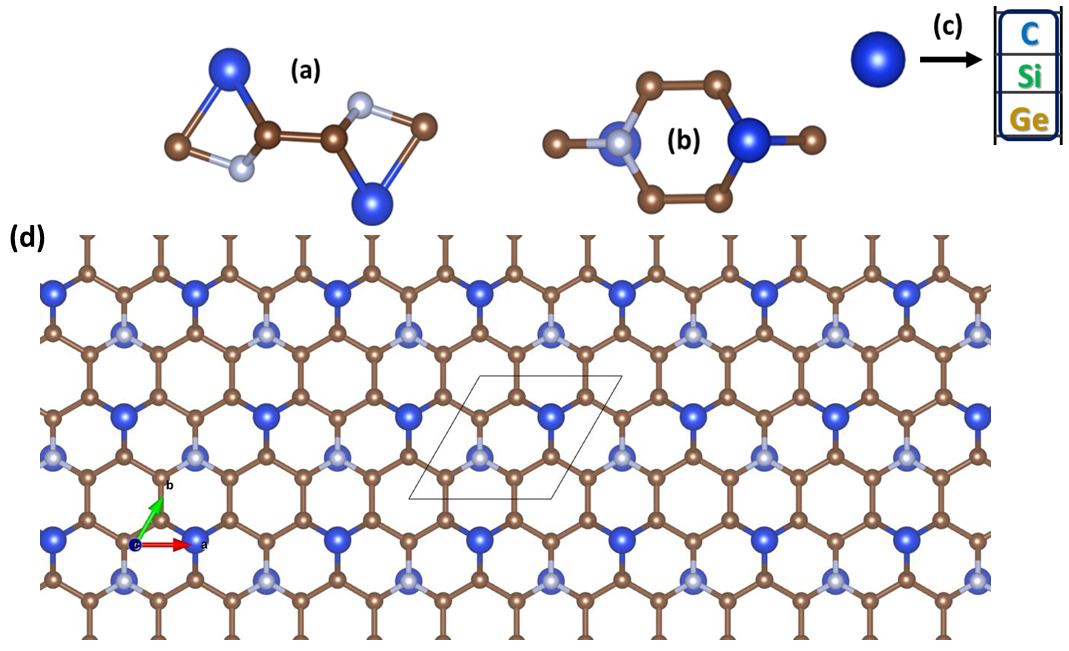}
    \caption{\textbf{Relaxed unit cell of the family of semi-metallic dumbbell C$_3$NX system.} Here figure (a) shows the side view, (b) the top view and (c) the adatoms responsible for dumbbell formation. Figure-(d) shows the full 2D sheet which is periodic in both x and y directions. The C (brown), N (grey) and X (blue) are the carbon, nitrogen and adatoms (C, Si and Ge) respectively.} 
    \label{Structure}
\end{figure*}
\begin{figure*}[ht]
    \centering
    \includegraphics[scale=0.6]{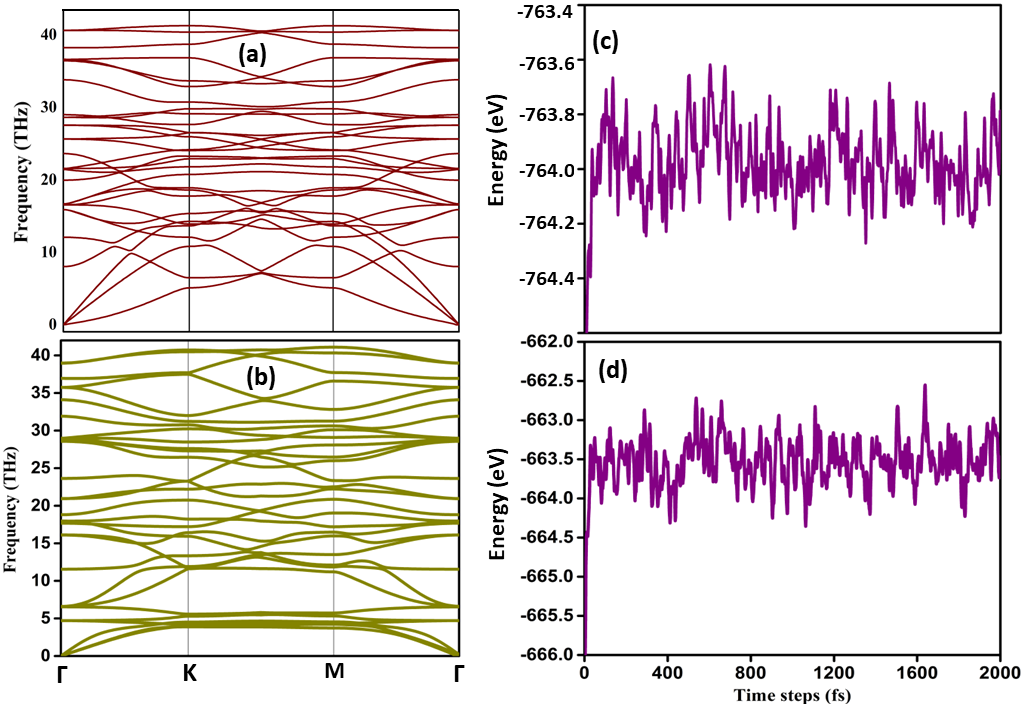}
    \caption{\textbf{Phonon dispersion curve and ab-initio molecular dynamics (AIMD) plot for semi-metallic dumbbell C$_4$N and C$_3$NGe.} Here figures (a) and (c) show the phonon dispersion and the AIMD plot for DB C$_4$N whereas figures (b) and (d) are for DB C$_3$NGe. The AIMD plots for these materials are examined and found in the canonical ensemble (NVT) while exposed to a 500 K and 300 K heat bath respectively for 2000 fs (with a time step of 1 fs) without any structural breakage.}
    \label{MD_Ph}
\end{figure*}

The concerned family of structures DB-C$_3$NX is obtained through the adsorption of some special adatoms (B, C, Si, Ge, P and As) from group-III, IV and V of the periodic table on the recently synthesized monolayer of C$_3$N or the two-dimensional polyaniline \cite{yang2017c3n,mahmood2016two}. The host atom N gets pushed by the two adatoms (X) on the top (and bottom) and forms a distinct buckled dumbbell-like morphology (Fig.~\ref{Structure}). Interestingly, only those elements participate in forming this unique structural arrangement whose electronegativities are comparable to that of carbon atoms. As far as structural symmetries are concerned, all these materials possess inversion symmetry about the center of the hexagon as opposed to the parent C$_3$N which had both reflection and inversion present. The C/N atoms in the C$_3$N monolayer have sp$^2$ hybridization similar to the atoms in graphene whereas in our case all the X(adatoms)/N atoms in the DB C$_3$NX monolayers find their ground state in sp$^3$ hybridization, resembling the likes of diamond structure. Magnetic calculations also reveal the non-magnetic ground state of these dumbbells. The bond lengths of these systems are also very different from what monolayer C$_3$N and graphene possess. All the relaxed optimized lattice constants, bond lengths between different atoms, buckling length and the corresponding cohesive energies of the semi-metallic dumbbell structures have been shown in our previous work \cite{jana2023spontaneous}. The negative signs and relatively comparable energy values with the synthesized C$_3$N monolayer depict decent energetic stability and feasibility of these materials.\\ 
The recent experimental realization of dumbbell (DB) silicene on Ag surface \cite{leoni2021demonstration} added to the stably grown DB phases of germanene on Pt and Au surfaces \cite{özçelik2014new}, this newer 2D morphology is attracting peers in the research community. All these dumbbell geometries including the DB stanene \cite{zhang2016quantum} are found to be energetically much more favorable than their low-buckled formations. Incorporating the hexagonal boron nitride (hBN) as a suitable substrate \cite{britnell2012field}, structures like these exhibit strong traces of Quantum anomalous hall effect. All these DB structures are found to be formed without any energy barrier once Si, Ge and Sn adatoms are placed on silicene, germanene or stanene \cite{ozcelik2015adsorption,ozccelik2013self} which speaks volumes about their synthetic plausibility. So to project the high synthetic feasibility of this family DB C$_3$NX (X = C, Si, Ge) we have first computed the formation energy of viz. the DB C$_4$N monolayers using the relation:
\begin{equation}
    \Delta E = \frac{E_{C_{4}N}-E_{C_{3}N}-nE_{C}}{n},
\label{form}
\end{equation}
where E$_{C_{4}N}$, E$_{C_{3}N}$ and E$_{C}$ are the energies corresponding to the DB C$_4$N, C$_3$N and the supplied carbon atoms whereas `n' is the number of atoms adsorbed. This E$_C$ is highly dependent on the form of supplied carbon resources. Using graphene as the carbon resource, the positive formation energy for DB C$_4$N comes out to be about 4.81 eV/carbon atom. This value resembles the energetics of depositing C atoms in graphitic C$_3$N$_4$ \cite{yang2015electron}. It is to note that, the carbon self-doped g-C$_3$N$_4$ has been achieved experimentally, which can enhance the proportion of C and N atoms \cite{dong2012carbon,wang2022synthesis}. Also using single C atoms as the carbon resource, the formation energy of the DB C$_4$N turns out to be 3.45/carbon atom. This is of the same order of magnitude as the adsorbed C atoms on graphene. The carbon atoms and their atomic chains are already been experimentally realized on the graphene surface using the transmission electron microscope (TEM) in ambient conditions \cite{ataca2011adsorption,meyer2008imaging}. Even though the synthetic route of the substitution pattern is quite different from that of the adsorption, we expect that the group-IV elements like C, Si etc. can be adsorbed on graphene-like C$_3$N forming DB C$_4$N, DB C$_3$NSi etc. since under adatom-rich environment, it's almost barrierless as these earth abundant sources are supplied \cite{yang2015electron,lourencco2016theoretical}. To properly establish the aspects of both dynamical and thermal stability, as a representative of this family of systems, we have added the phonon dispersion curves and the ab-initio molecular dynamics for DB C$_4$N and C$_3$NGe here respectively. In Fig.~\ref{MD_Ph}(a,b) the absence of any negative vibrational mode defines the dynamical stability of these materials. The variation of total energy with time in Fig.~\ref{MD_Ph}(c,d) depicts thermal stability for these structures while exposed to a 500 K heat bath for 2000 fs without any bond breakage.
The synthesis of any film, regardless of the dimensionality, typically requires a substrate. Proper growth of these C$_3$N-based dumbbells can be catered by the widely known approach which is to form a vdW heterostructure. The hBN sheets possess a large bandgap and higher dielectric constant which enabled it as a worthy substrate to grow graphene and other 2D materials \cite{britnell2012field}. Forming a vertical heterojunction with graphene-like C$_3$N and hexagonal BN \cite{mondal2024c3n} and depositing additional C, Si atoms on the surface can lead to the formation of DB C$_3$NX on hBN. For example, a unit cell of DB C$_3$NX (`Si' here) can be placed between two 2$\times$2 hBN - forming a vdW heterostructure with a low lattice mismatch of only 1\%. Figure-S1 in the supplementary material reveals the fully relaxed trilayer encapsulation of DB C$_3$NSi with boron nitride (BN), accompanied by its associated band structure. Remarkably, the Dirac cone characteristic of free-standing DB C$_3$NSi is preserved around the Fermi energy when encapsulated in BN, owing to the retention of inversion symmetry. Therefore, when DB C$_3$NX is sandwiched between hBN layers, it showcases a topologically nontrivial gap induced by spin-orbit coupling. This strikingly contrasts with the scenario where DB C$_3$NX rests on a single BN layer, which results in a 60 meV bandgap around the Fermi level due to the broken inversion symmetry, as depicted in Figure~S2.

\begin{figure*}[ht]
    \centering
    \includegraphics[scale=0.52]{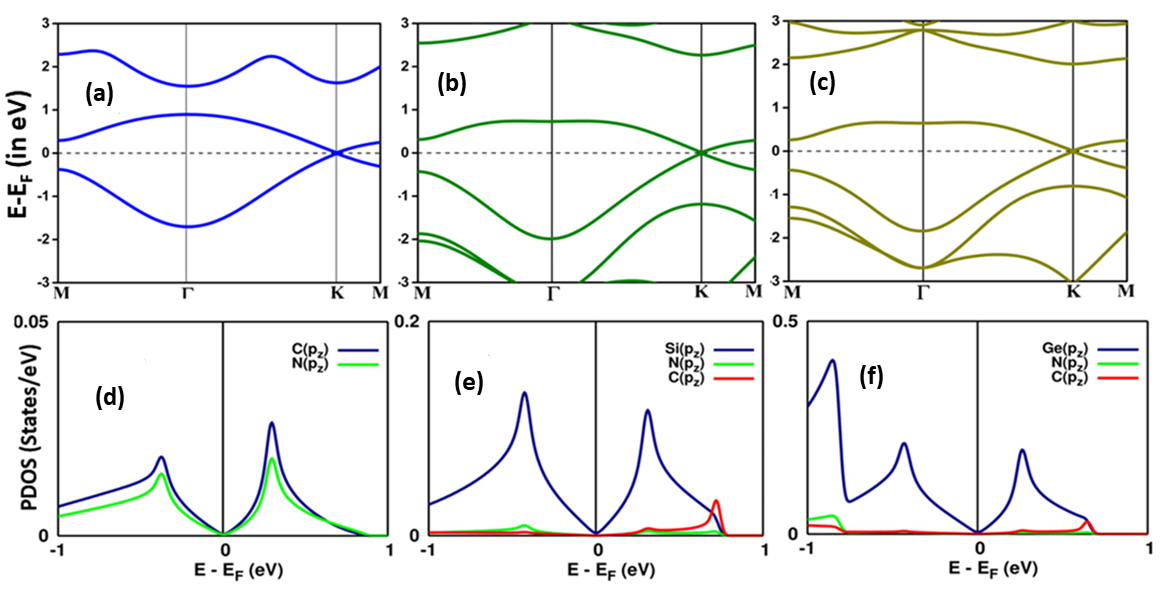}
    \caption{\textbf{Electronic structure of the family of semi-metallic dumbbell C$_3$NX system.} Here figure (a), (b) and (c) shows the existence of the Dirac cone in all three DB C$_4$N, C$_3$NSi and C$_3$NGe at high symmetry `K' point. Figure-(d,e,f) shows the atom- (and corresponding orbital) resolved density of states for all three dumbbell formations respectively.} 
    \label{Band}
\end{figure*}

From an experimental point of view, modulation via strain is also quite important for the fabrication of such 2D monolayers. In our earlier work \cite{jana2023spontaneous}, we have performed both the structural modification and band tuning through the uni- and bi-axial strains. In both cases, the Dirac points are robust up to about (2-4)\% strain value above which they get opened up to bandgaps of meV order around the high symmetry `K’ point. As per reports, one-third-hydrogenated graphene (C$_3$H) single-crystal is synthesized experimentally on Ru (0001) surface by Chen et al. \cite{chen2018fabrication} where the Dirac point is preserved. So a similar hydrogenation of these reported dumbbell carbon nitrides on proposed substrates like Ag or hBN can be fruitful. So when these monolayers are grown on the substrates proposed above, the usual lattice constant dissimilarities can't affect their electronic nature easily.\\
Electronically, dumbbells formed by the group-IV elements (C, Si and Ge) portray the presence of Dirac cones at the high-symmetry point `K' as shown in Fig.~\ref{Band}(a,b,c). This emergence of Dirac cone is obvious due to the simultaneous existence of both spatial inversion and time-reversal symmetry \cite{wang2015rare}. 
As opposed to single-layer carbon-nitride systems like g-C$_6$N$_6$ \cite{wang2014topological} and C$_3$N, these three DB C$_3$NX structures have the Dirac-like crossing exactly at the Fermi level, analogous to holey C$_x$N$_y$ monolayers like C$_7$N$_3$, C$_{10}$N$_3$ etc \cite{tan2021dirac} which is better for experiments and applications. Although the X/N atoms are sp$^3$ hybridized, these dumbbells do not mix up other orbitals. Instead, near the Fermi energy level, only the p$_z$ orbitals of the adatoms seem to be responsible for the formation of the Dirac cones. This feature is quite enthralling since other Dirac semimetals like monolayer phagraphene p$_z$ orbitals of 8 atoms out of the 20 carbon atoms in the unit cell contribute in the Dirac-cone formation \cite{ghosh2022intriguing}. The atom-resolved projected density-of-states distinctively shows this feature in Fig.~\ref{Band}(d,e,f). This also indicates that if we eliminate all the other insignificant atoms and keep only the contributing adatoms - that low-energy structure should surely reproduce the Dirac cone at the Fermi energy level. A detailed analysis of the structure and the electronic nature of this family of materials has been done in our earlier work \cite{jana2023spontaneous}. 
\section{Topological phase transition in C$_3$NX sheet}
\begin{figure}
\centering
\includegraphics[scale=0.30]{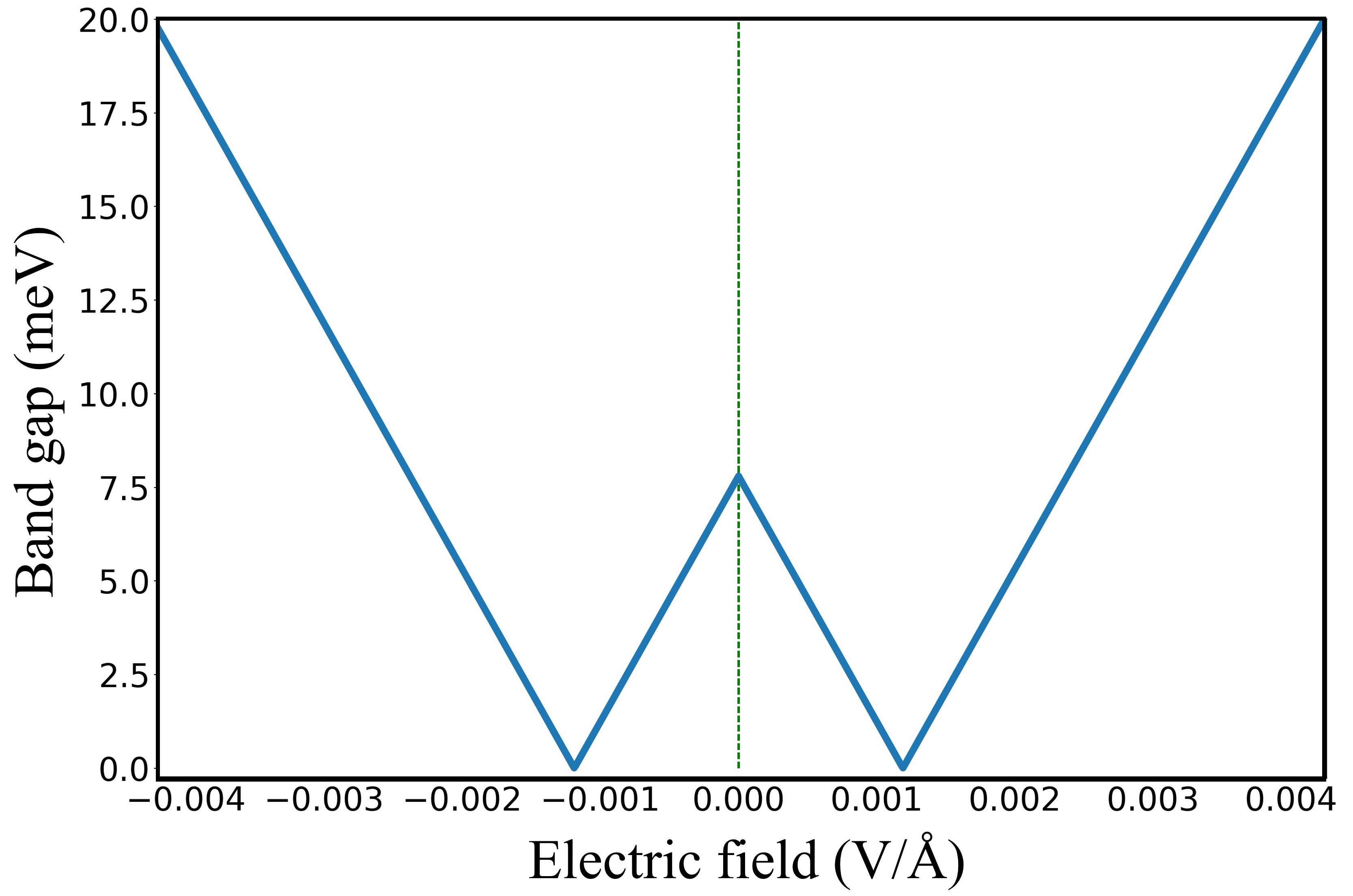}
\caption{\textbf{The electrical tunability of the band gap and the indication of a possible topological phase transition in C$_3$NX systems (in this case, X=Si) is done through the TB model.} The band gap exhibits a linear variation with the electric field, decreasing before reaching a critical value. Beyond this value, the band gap increases. At the critical electric field strength, the system behaves as a semimetal with zero band gap value. } 
\label{fig:bandgap}
\end{figure}
Delving into the intricacies of these systems, we present the tight-binding model Hamiltonian tailored explicitly for the effective low-energy lattice version of the C$_3$NX systems. We note that our low-energy lattice model, built upon the consideration of effective hopping between the adatoms of dumbbell-like structures, is well-equipped to explore the essential physics of the systems near the Fermi level. This is because of the predominant contribution of the adatom orbitals near the Fermi level of all the systems considered in this work. Despite the shared essence of their fundamental geometric attributes, these lattices differ in some essential parameters, such as bond lengths ($d$), hopping integrals ($t$), and lattice parameters ($a$). Besides, the SOC strength ($\lambda_{SO}$) and buckling height ($l$) between adjacent `X' atomic sites depend sensitively on the atomic number of constituent elements, exhibiting quantitatively distinct characteristics across different systems. The presence of substantial buckling in the system allows a transverse electric field ($E_{z}$) to impart different mass terms (Semenoff-like) \cite{bandyopadhyay2020topology,sarkar2023non} to the neighboring atoms of the effective low-energy structure. The Semenoff mass essentially results in the breakdown of inversion symmetry within the system,
leading to the emergence of non-zero Berry curvature even in the presence of time-reversal symmetry. The generalized TB Hamiltonian \cite{kane2005quantum} for our effective buckled honeycomb lattices, subject to an external transverse electric field, can be expressed as:
\begin{align*}
     \hat{H} = & - t \sum_{\langle ij\rangle,\sigma} c_{i,\sigma}^{\dagger}c_{j,\sigma} + i \frac{\lambda_{SO}}{3\sqrt{3}} \sum_{\langle\langle ij\rangle\rangle ,\sigma} \sigma \zeta_{ij} c_{i,\sigma}^{\dagger}c_{j,\sigma} \nonumber
      -l \sum_{i,\sigma} \nu_{i} E_{z} c_{i,\sigma}^{\dagger}c_{i,\sigma}.
\label{eq:ham}
\end{align*}
In the presented theoretical framework, the symbols $\langle ij\rangle$ and $\langle\langle ij\rangle\rangle$ denote the processes of electron hopping between sites $i$ and $j$ within a lattice structure, encompassing interactions with nearest and next-nearest neighbors. The spin degrees of freedom, denoted by the symbol $\sigma$, manifest as either $\uparrow$ ($+1$) or $\downarrow$ ($-1$), representing the intrinsic spin states. Additionally, the system incorporates the influence of SOC and staggered sublattice potential denoted by $\zeta_{ij}$ (taking values of $\pm 1$) and $\nu$ (with values of $\pm 1$) respectively. These parameters explicitly hinge upon the direction (clockwise or anticlockwise) of the hopping process and the specific sublattice involved. It is essential to mention that the tight-binding parameters governing Eq.~\ref{eq:ham} exhibit variations across diverse systems under consideration. Table \ref{tab:tabls} provides a comparison of these parameters for the dumbbell-like structures in relation to Xenes (X= C, Si, Ge, and Sn).
\begin{table}
\centering
\caption{The structural and tight binding model parameters for dumbbell-like structures are systematically compared with those of graphene, silicene, germanene, and stanene. In this analysis, the symbols $a$, $t$, $\lambda_{so}$, and $l$ represent fundamental properties such as the lattice parameter, hopping integral between adjacent atoms, intrinsic spin-orbit coupling, and buckling heights, respectively.}
    \setlength{\tabcolsep}{0.5em}
    \def\arraystretch{1.2}%
    \begin{tabular}{lllll}
    \hline
    \hline
     Systems & $a$ (\AA{}) & $t$ (eV)  & $\lambda_{SO}$ (meV) & $l$ (\AA{}) \\
     \hline
     \hline
    DB C$_{4}$N & 4.76 & 2.8 & 10$^{-3}$ & 2.01 \\
    DB C$_{3}$NSi & 4.89 & 0.21 & 3.9 & 3.28 \\
    DB C$_{3}$NGe & 4.90 & 0.232 & 43 & 3.62 \\
    Graphene & 2.46 & 2.8 & 10$^{-3}$ & 0.00 \\
    Silicene & 3.86 & 1.6 & 3.9 & 0.23 \\
    Germanene & 4.02 & 1.3 & 43 & 0.33 \\
    Stanene & 4.70 & 1.3 & 100 & 0.40 \\
     \hline
     \end{tabular}
     \label{tab:tabls}
\end{table}
Notably, the negligible spin-orbit coupling and planar geometry of graphene preclude the manifestation of an electrically tunable quantum spin Hall state in this system. On the contrary, all of these dumbbell-like structures are buckled and exhibit a band gap in the unperturbed case. Moreover, it is worth noting that the buckling heights of these dumbbell-like structures are significantly large compared to any of the Xenes (X = C, Si, Ge, Sn). Therefore, the Hamiltonian equation (Eq.~\ref{eq:ham}) elucidates the advantage of modulating band gaps by external electric field modulation of these dumbbell-like structures. The physics in the vicinity of the Fermi energy is aptly characterized by the Dirac theory, a framework analogous to that applied in graphene. The presence of electric field ($E_z$) modifies the electronic band dispersion of the buckled dumbbell-like structures by the relation $\mathcal{E}(k) = \pm \sqrt{(\hbar v_F k)^2 + (l E_z - \eta \sigma \lambda_{so})^2} $, where $\eta = \pm 1$ represents valley index for K and K$^{\prime}$ points. In the absence of any external electric field, the buckled dumbbell-like structures inherently possess a SOC-induced band gap, $\Delta \mathcal{E} = 2 \lambda_{SO}$, as also obtained from our first-principles methods.
\begin{figure*}[ht]
    \centering
    \includegraphics[scale=0.48]{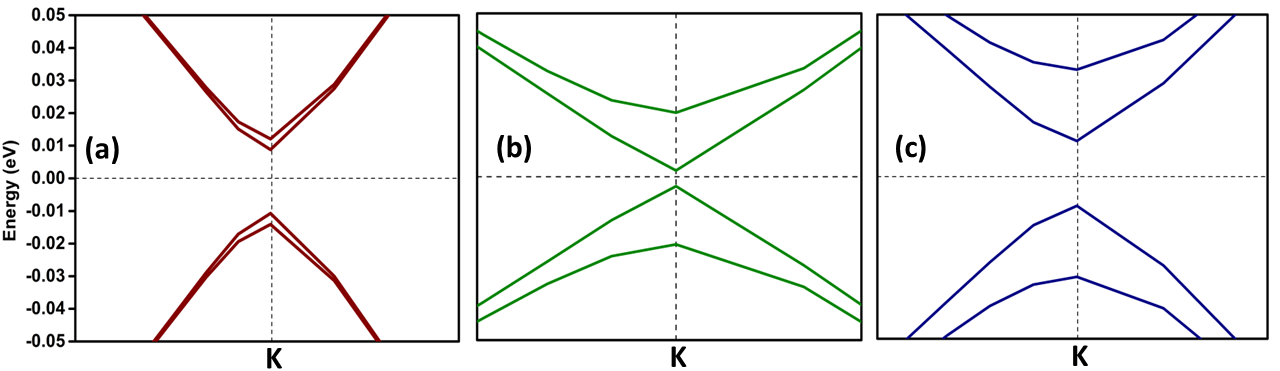}
    \caption{\textbf{Variation of band gap of DB C$_3$NGe system with transverse electric field.} Here (a) E$<$ E$_c$ (b) E $\approx$ E$_c$, and (c) E $>$ E$_c$ is computed using DFT. The band gap first decreases with the electric field which causes a topological phase transition and then increases again to a normal insulating phase.}
    \label{Band_field}
\end{figure*}
\begin{figure*}[!bhtp]
\centering
\includegraphics[scale=0.56]{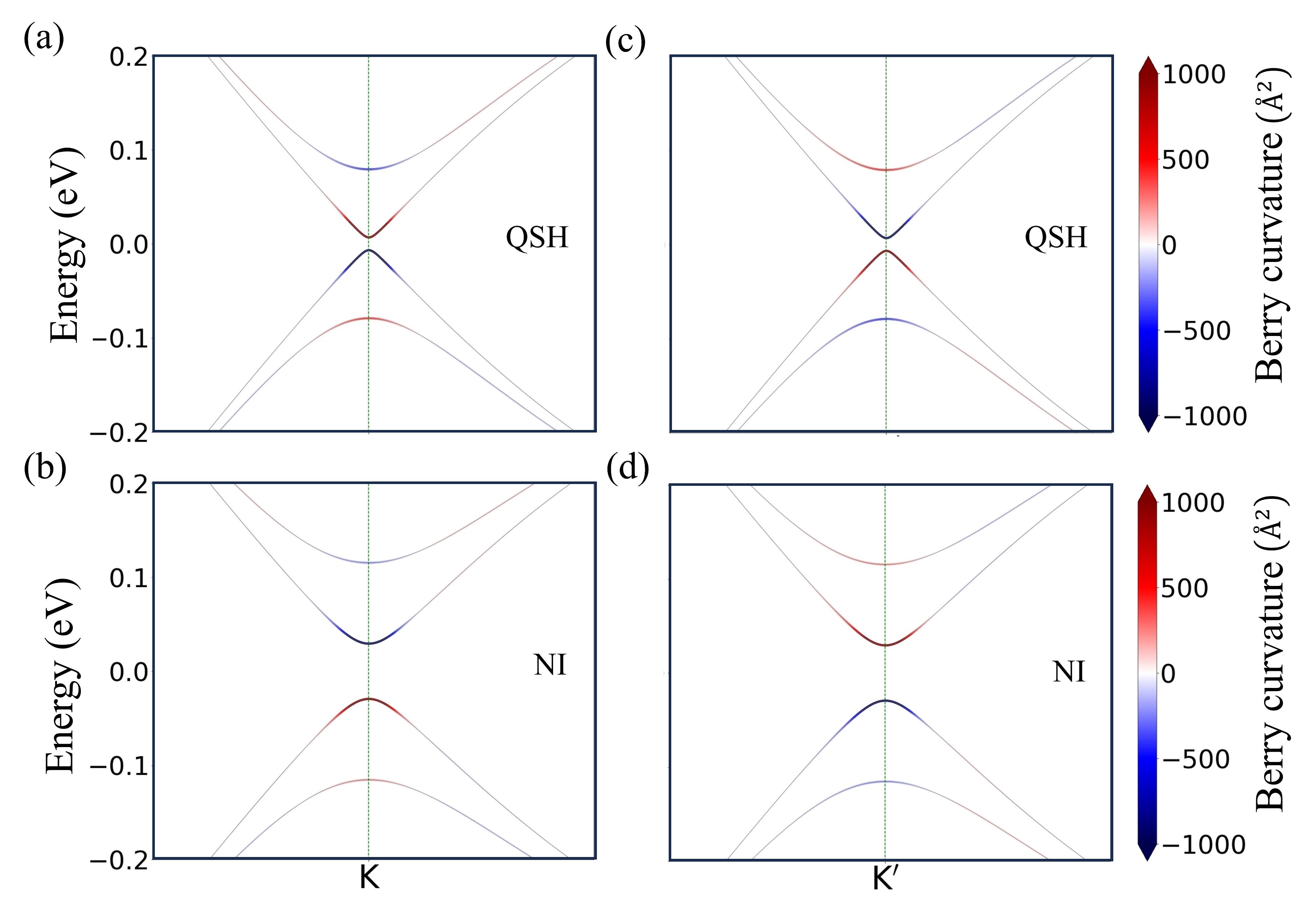}
\caption{\textbf{Berry curvature distribution across four spin bands near the Fermi level under varying electric fields in C$_{3}$NGe which is computed via realistic tight-binding method.} Panels (a) and (b) illustrate this near $\mathcal{K}$, with field strengths of 0.01 V/\AA{} (nontrivial phase) and 0.02 V/\AA{} (trivial phase). Panels (c) and (d) show effects around $\mathcal{K}^{\prime}$ for the same field strengths. Electric fields cause spin degeneracy splitting, and inversion of Berry curvature in two bands signifies the topological phase transition from quantum spin Hall (QSH) to normal insulating (NI) phase. The color scale indicates Berry curvature magnitude and sign, with $\Omega_{z}$ as the sole non-zero component in 2D.}
\label{fig:berry_curvature}
\end{figure*}
The given expression of band gap indicates a decreasing SOC-induced band gap with electric field strength until it reaches zero, followed by the emergence of a linearly varying band gap with increasing electric field as depicted in Fig.~\ref{fig:bandgap}. In Fig.~\ref{fig:bandgap}, the particular example of C$_3$NSi exhibits characteristics of a QSH insulator in the unperturbed phase with a SOC-induced band gap of 7.8 meV. However, the application of an external electric field induces a reduction in the band gap, eventually leading to its closure at the critical electric field value of 0.0012 V/\AA{}. Beyond these critical points, a transformative shift transpires, causing C$_3$NBerrySi to exhibit properties of a normal insulator. The band gap in the normal insulating case increases linearly with the external electric field, as discussed earlier.
Here we note that, all three materials (C$_3$NX) are isostructural, differing only in the type of X atom (X = C, Si, Ge). The intrinsic electronic behavior of these systems is essentially similar, with primary differences in the vertical separation between neighboring X atoms and the SOC-induced band gap. Table 1 of our manuscript compares these properties. The strength of SOC ($\lambda_{so}$) is responsible for the band gap opening in the unperturbed state, rendering these systems topologically nontrivial. Conversely, the vertical distance separation ($l$) between neighboring X atoms induces an onsite potential difference in the presence of a transverse electric field, which tends to make the systems trivial. The competition between these two parameters drives the topological phase transition around a critical electric field strength of $\lambda_{so}/l$. Therefore, from this discussion, it is evident that systems with stronger SOC require a stronger electric field to induce the topological phase transition, while a larger vertical separation reduces the critical field value. In order to corroborate the field driven topological phase transition in C$_3$NX systems we have performed DFT calculations. 
It is evident from Fig.~\ref{Band_field} that the band gap at the $K$ point initially decreases with the application of an external transverse electric field, inducing a topological phase transition, and subsequently increases, forming a trivial band gap. Therefore, the above discussion presents a procedure for the controlled manipulation of the electronic properties of C$_3$NX systems through electrical means. Moreover, the modulation of the band gap suggests the potential for an exciting interplay among external electric fields, electronic band structures, and the topological phases across all C$_3$NX systems. To properly illustrate the nature of distinct band gap opened phases, i.e., before and after the gap closing value of the electric field strength, we have calculated the Berry curvature ($\Omega$) distribution \cite{bandyopadhyay2024non,bandyopadhyay2023berry,bandyopadhyay2022electrically} on the electronic bands near both the valleys $\mathcal{K}$ and $\mathcal{K}^{\prime}$ using Kubo formalism \cite{xiao2010berry} a given below.
%
\begin{equation}
    \Omega_z (\vec{k}) = 2 i \sum_{i \neq j} \frac{\mel{i}{\partial\hat{H}/\partial k_x}{j} \mel{j}{\partial\hat{H}/\partial k_y}{i}}{\left(\varepsilon_i - \varepsilon_j\right)^{2}},
\label{eq:bc}
\end{equation}
\begin{figure*}
\centering
\includegraphics[scale=0.54]{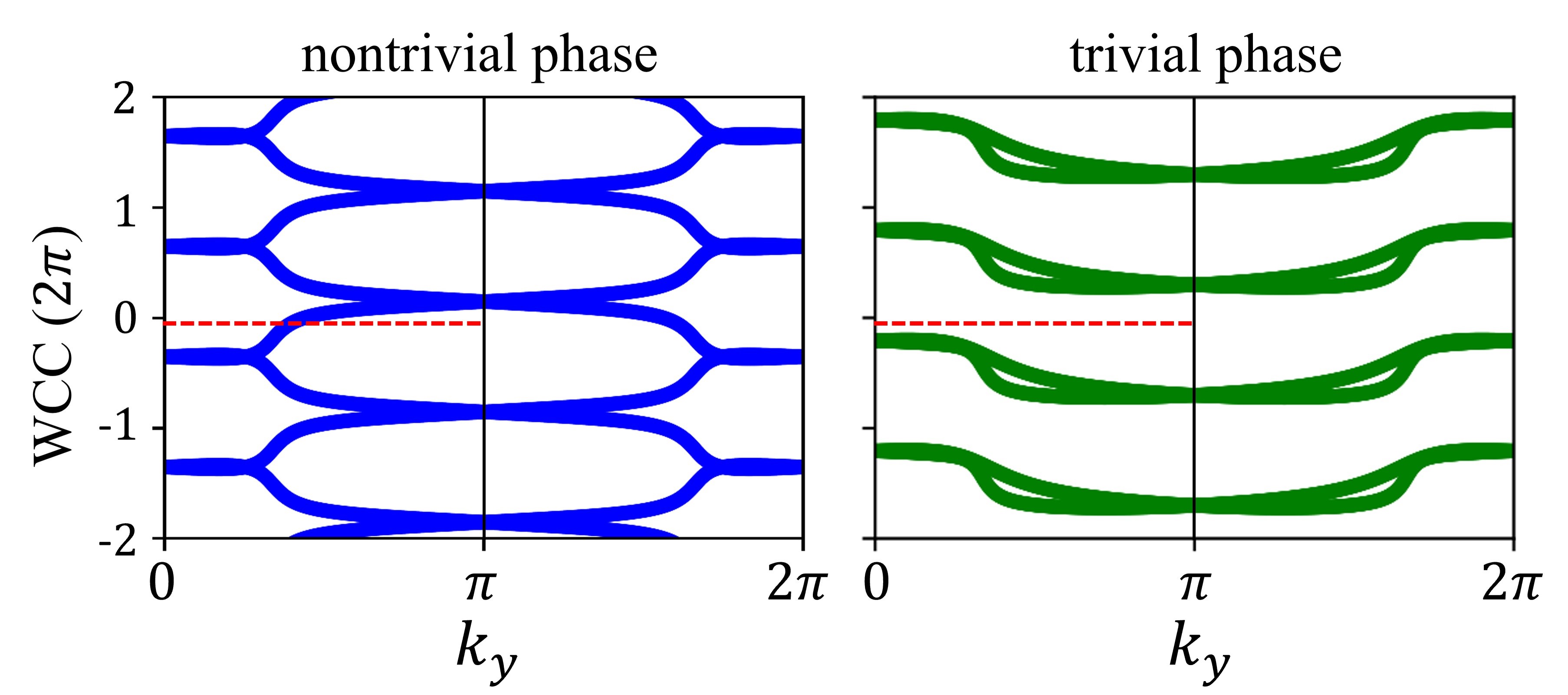}
\caption{\textbf{Wannier charge centers in nontrivial and trivial phases}. The Wannier charge centers (WCCs) are depicted for DB C$_{3}$NGe under distinct electric field strengths: (a) 0.01 V/\AA{} (representing the nontrivial phase) and (b) 0.02 V/\AA{} (indicating the trivial phase), evaluated through a realistic tight-binding method. In the former case, a single line parallel to $k_y$ axis intersects an odd quantity of charge centers within half of the Brillouin Zone (BZ), resulting in an odd $\mathbb{Z}_2$ invariant. This odd $\mathbb{Z}_2$ value substantiates the nontrivial topological nature of the material. Conversely, in the latter case, the WCC profile indicates an even number of intersection points, signifying an even $\mathbb{Z}_2$ invariant and confirming the trivial phase.} \label{fig:wcc}
\end{figure*}
We note that, in two-dimensional crystals only the out-of-plane component of Berry curvature ($\Omega_z$) is non-zero. The eigenenergies of the Hamiltonian $\hat{H}$, denoted as $\varepsilon_i$ and $\varepsilon_j$, correspond to the eigenstates $\ket{i}$ and $\ket{j}$, respectively. It is noteworthy that the Berry curvature of systems exhibiting time-reversal invariance possesses an intrinsic odd symmetry with respect to momentum. This is expressed as $\Theta^{\dagger} \Omega_z(-\vec{k}) \Theta = - \Omega_z(\vec{k})$, where $\Theta$ denotes the time-reversal operator. Moreover, Berry curvature plot in Fig.~\ref{fig:berry_curvature} reveals a clear band inversion around the critical electric field strength $\approx \lambda_{SO}/l = 0.012$ V/\AA{} for C$_{3}$NGe system. This band inversion indicates the topological phase transition in the C$_3$NGe system about the critical electric field 0.012 V/\AA{}. 
\begin{figure}
\centering
\includegraphics[scale=0.20]{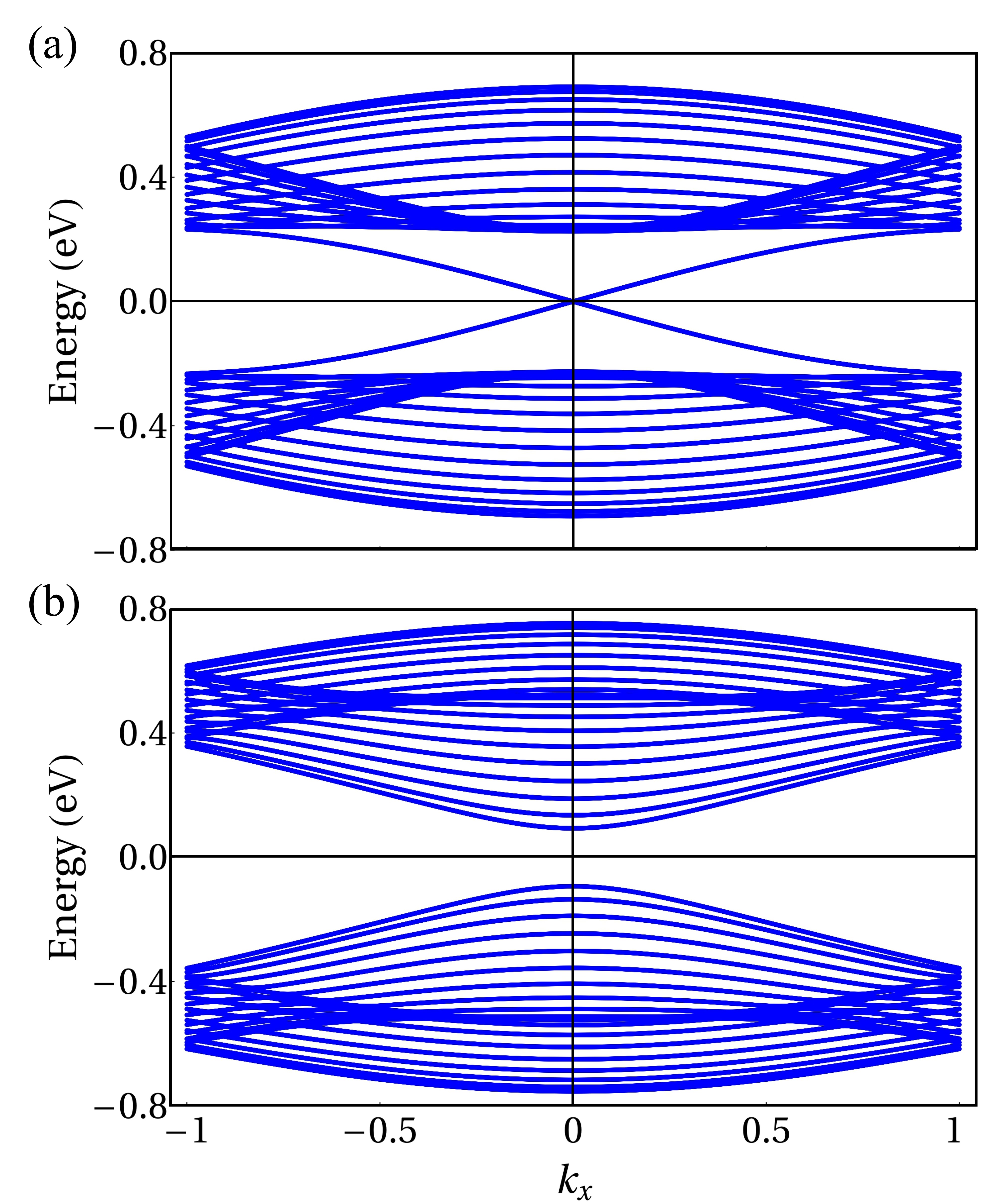}
\caption{\textbf{The band structure of the armchair edge nanoribbon formed by the X (= Ge) atoms is computed using realistic tight binding method.} (a) In the absence of an electric field, the C$_3$NGe nanoribbon exhibits protected metallic edge states, which corresponds to the bulk topological invariant $\mathbb{Z}_2 =1$. (b) In the presence of a perpendicular electric field beyond the critical value (here, the mass term = 0.3 eV), the bulk topological invariant becomes trivial  $\mathbb{Z}_2 =0$, and consequently, there are no edge states. Therefore, these two phases are topologically distinct. Here, 20 hexagonal plaquettes are considered along the non-periodic direction.}\label{fig:edgestaes}
\end{figure}
It is noteworthy that the same underlying physics manifests in the C$_{3}$NSi system, where the semimetallic phase is achieved at a critical electric field strength of 0.0012 V/\AA{} about which system undergoes a topological phase transition. To establish the presence of distinct topological phases in C$_{3}$NX systems more strongly, we have calculated the $\mathbb{Z}_2$ invariant using Wannier charge centers (WCC) \cite{yu2011equivalent, soluyanov2011computing}.
In particular, we have calculated Berry phases around the BZ in the k$_x$ direction, which can be interpreted as the 1D hybrid Wannier centers in the x-direction and plot results as a function of k$_y$. The formulation of the $Z_2$ invariant from the WCC, introduced by Fu and Kane \cite{fu2006time}, expresses the $Z_2$ invariant as an obstruction of the $U(1)$ Berry phase gauge field within half of the BZ. This method offers convenience because it resembles the Chern number formula for quantum Hall states, as established by Thouless et al. \cite{thouless1982quantized}. Later, Yu et al. \cite{yu2011equivalent} offered an alternative proof demonstrating the relationship between the Wilson loop approach and the $Z_2$ invariant via the obstruction formula. The occupied Bloch bands can be written down using the gauge choice $|n,-k> = \mathcal{T}_{n,m} T(|m,k>)$. It is needless to mention that the operator $\mathcal{T} : \mathcal{T}^2 = -1 $ represents an anti-symmetric matrix. Using the gauge choice mentioned above, the $Z_2$ invariant for any time reversal invariant Hamiltonian can be expressed as follows:
\begin{equation}
 \Delta  = \frac{1}{2\pi} \left( \int_{0}^{\pi} dk_y \partial_{k_y} \Phi(k_y) - [\Phi(\pi) - \Phi(0)] \right) \, \text{mod} \, 2.
 \label{eq:z2exp}
\end{equation}
In the Eq.~\ref{eq:z2exp}, $\Phi$ is the $U(1)$ Wilson loop which can be determined form $U(1)$ part of the Berry connection ($a_i$) as $\Phi(k_y) = \oint_{-\pi}^{\pi} dk_x \, a_x(k_x,k_y)$. Here, we note that the wave functions in the lower half of the BZ, defined by \( k_y \in [-\pi, 0] \), can essentially be determined by those in the upper half of the BZ, denoted by \( \tau_{1/2} \). Considering wavefunctions to be continuous and well defined in $\tau_{1/2}$, it is evident from Eq.~\ref{eq:z2exp} that the $Z_2$ invariant is either 0 or any other arbitrary integer. For a particular gauge choice \cite{yu2011equivalent}, $|n,k> \rightarrow e^{i \phi_k} |nk>$ one can write,  
\begin{equation}
 \Delta  = \frac{1}{2\pi} \left( \sum_n \int_{0}^{\pi} dk_y \partial_{k_y} \phi_{n}(k_y) - \sum_{n}[\phi_{n}(\pi) - \phi_{n}(0)] \right) \, \text{mod} \, 2.
 \label{eq:z2exp}
\end{equation}
Considering $\phi_{n}[0]$ and $\phi_{n}[\pi]$ within [0, 2 $\pi$) we can write $\int_{0}^{\pi} dk_y \frac{\partial \phi_n(k_y)}{\partial k_y}  = \phi_n(\pi) - \phi_n(0) + 2\pi M_n$, where $M_n$ serves as the winding number. In principle, winding number in this case corresponds to the number of times $\pi_n$ crosses the horizontal line ($\phi_n = const$) of Wannier charge center plot in the half BZ. The above discussion leads us to derive a simple definition of the $Z_2$ index, ie., $\Delta = \sum_{n} M_{n}$ mod 2.
In Fig.\ref{fig:wcc}(a), it is evident that a linear trajectory parallel to $k_y$ intersects the WCC of C$_3$NGe an odd number of times in half of the Brillouin Zone under an electric field value 0.01 V/\AA{}, which is below the critical field. This odd number of intersections essentially establishes the nontrivial insulating behaviour with the topological index $\mathbb{Z}_2=1$. On the contrary, under a 0.02 V/\AA{} electric field (above the critical field), the same trajectory exhibits an even number of intersections showing the trivial topological index $\mathbb{Z}_2=0$, as depicted in Fig.\ref{fig:wcc}(b). The $\mathbb{Z}_2$ invariant conclusively establishes that the applied external electric field induces a topological phase transition in C$_{3}$NGe, from a nontrivial to a trivial one. Similar outcomes have been observed in other structures like dumbbell C$_3$NSi. However, due to the significantly low strength of the SOC in the C$_{4}$N system, the emergence of the quantum spin Hall state is suppressed. In order to corroborate the electric field-driven topological phase in the two dimensional DB C$_3$NX systems, we have further calculated the topological edge states in the armchair nanoribbon geometry with 20 hexagonal plaquettes along the non-periodic direction. We note that, the bulk-boundary correspondence is a key principle in topological insulators, linking bulk properties with edge or surface states. A non-trivial bulk topological invariant $\mathbb{Z}_2$ essentially ensures the presence of robust, gapless states at the boundaries. These states are protected by the non-trivial bulk topological index $\mathbb{Z}_2$=1, making them immune to local perturbations that do not close the bulk energy gap and cause a topological phase transition. In the present case of two-dimensional topological insulators, the insulating bulk (non-trivial phase) hosts helical edge states where electron spin is locked to the direction of motion. This results in robust conduction channels resistant to backscattering and non-magnetic impurities that preserve time-reversal symmetry \cite{li2023built,wang2022quantum,li2017topological}. In order to further support the topological insulating properties of pristine C$_3$NX, we have provided the topologically protected edge states, given in Fig.~\ref{fig:edgestaes}(a) corresponding to the nontrivial topological invariant $\mathbb{Z}_2$=1 for the 2D sheet. Furthermore, we have demonstrated that these conducting edge states vanish beyond the critical value of the inversion symmetry-breaking perpendicular electric field [see, Fig.~\ref{fig:edgestaes}(b)], at which point the bulk topological invariant becomes $\mathbb{Z}_2=0$.
\section{C$_3$NX nanoribbon}
In this section, we shall investigate the electronic characteristics of the DB C$_3$NX nanoribbon and particularly focus on the potential for facilitating topological phase transitions even in the C$_{4}$N structure.
\begin{figure*}[!bhtp]
\centering
\includegraphics[scale=0.5]{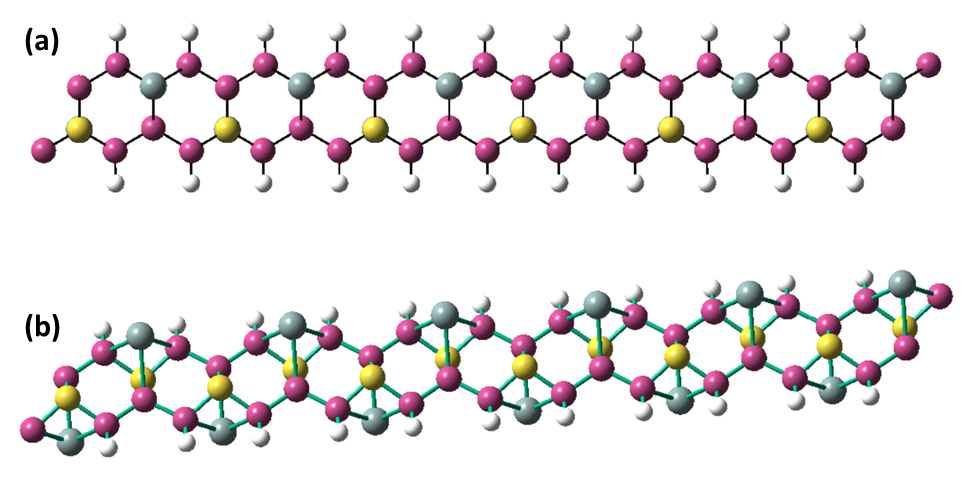}
\caption{\textbf{Top and tilted side view of the relaxed edge-passivated quasi-1D or nanoribbon form of DB C$_3$NX structure}. The red, yellow, grey and white atoms represent carbon (C), nitrogen (N), adatoms (X = C, Si, Ge) and hydrogen (H) respectively. This is a 6$\times1\times$1 unit of the periodic one-dimensional chain of these concerned systems.}
\label{Ribbon_pass}
\end{figure*}
We start from the demonstration of the quasi one-dimensional version of the original network. The Fig.~\ref{Ribbon_pass}(a,b) here depicts the fully relaxed hydrogen-passivated quasi-1D formation of DB C$_3$NX which is basically a 6$\times1\times$1 block of its infinite one-dimensional network with the lattice periodicity only along crystallographic $\vec{a}$-axis of the original sheet depicted in Fig.~\ref{Structure}(d). As we can see from the Fig.~\ref{Ribbon_pass} that the edge type of the nanoribbon is zigzag and it has the width of a single hexagonal plaquette and the edges are terminated by hydrogen atoms in order to ensure the stability. As an example, the dynamical and thermal stability is checked and established for quasi-1D DB C$_3$NX systems and given for X=Si in Figure-S2 of the supplementary material. Therefore, the unit cell of the nanoribbons consists of C, N, X and H atoms, however, the projected density of states calculation using density functional theory reveals that the states near the Fermi level are solely contributed by the p$_z$ orbital of X atom.
The ribbon can be demonstrated by the standard tight-binding Hamiltonian, written in
the Wannier basis for the spinless non-interacting fermions, viz.,
\begin{equation}
H=\sum_{\langle jk \rangle} [t_{jk} a_{j}^{\dagger} a_{k} + h.c.] + \sum_{j} \epsilon_{j} a_{j}^{\dagger} a_{j}
\label{hamil}
\end{equation}
The off-diagonal term is the overlap integral between the nearest neighboring nodes that carries the kinetic signature and the diagonal term describes the potential information of the respective atomic site location. There are three different kinds of atomic sites. Depending on the local environment we may assign different potential for different quantum dot locations. However, the PDOS and band spectra of the systems indicate that we can examine the spectral landscape provided by the ribbon geometry by considering uniform on-site energies of every kind of atomic site. The cost we need to pay is a shift of Fermi energy compared to the charge neutrality of the model because of the presence of nitrogen in the systems. For further numerical evaluation, we set $\epsilon=0$ for
all the nodes. 
We will try to evaluate the spectral information by the real space decimation scheme \cite{mukherjee2023application}.
It is needless to say that the decimation scheme is no longer dependent on the choice of numerical values of the parameters of the Hamiltonian.
Now, the discretized form of the Schr\"{o}dinger's equation
(equivalently termed as \textit{difference equation}) can be written as
\begin{equation}
(E-\epsilon_{n}) \psi_{n} = \sum_{m} t_{nm} \psi_{m}
\label{diff}
\end{equation}
The summation on the right-hand side of the above equation includes the contribution of
all the nearest neighbor atomic sites. The longer wavelength fluctuation is not considered here.
To acquire an idea about the general spectral feature we will exploit  Eq.~\eqref{diff} and try to obtain an analytical overview of the band diagram. As the underlying lattice has translational invariance one should expect the obvious existence of Bloch states in the band dispersion.
For any such resonant mode, the system becomes completely transparent to the
incoming excitation.
By exploiting the real space decimation formalism, we can eliminate the wave function amplitudes of an appropriate subset of atomic sites in terms of the other surviving nodes. This transformed structure with the corresponding effective parameters of the Hamiltonian will now be useful for the numerical evaluation of the density of eigenstates (DOS) and the electronic transmission probability as a function of the energy of the incoming projectile for the entire ribbon. Here we should mention that the decimation scheme, mentioned above, is nothing but a mathematical artifact and the original network remains intact. Therefore, the analytical workout does not have the chance of destroying the overall spectral nature and hence we will not lose any serious information, physics-wise.
\begin{figure*}
\centering
\includegraphics[clip,width=14 cm,angle=0]{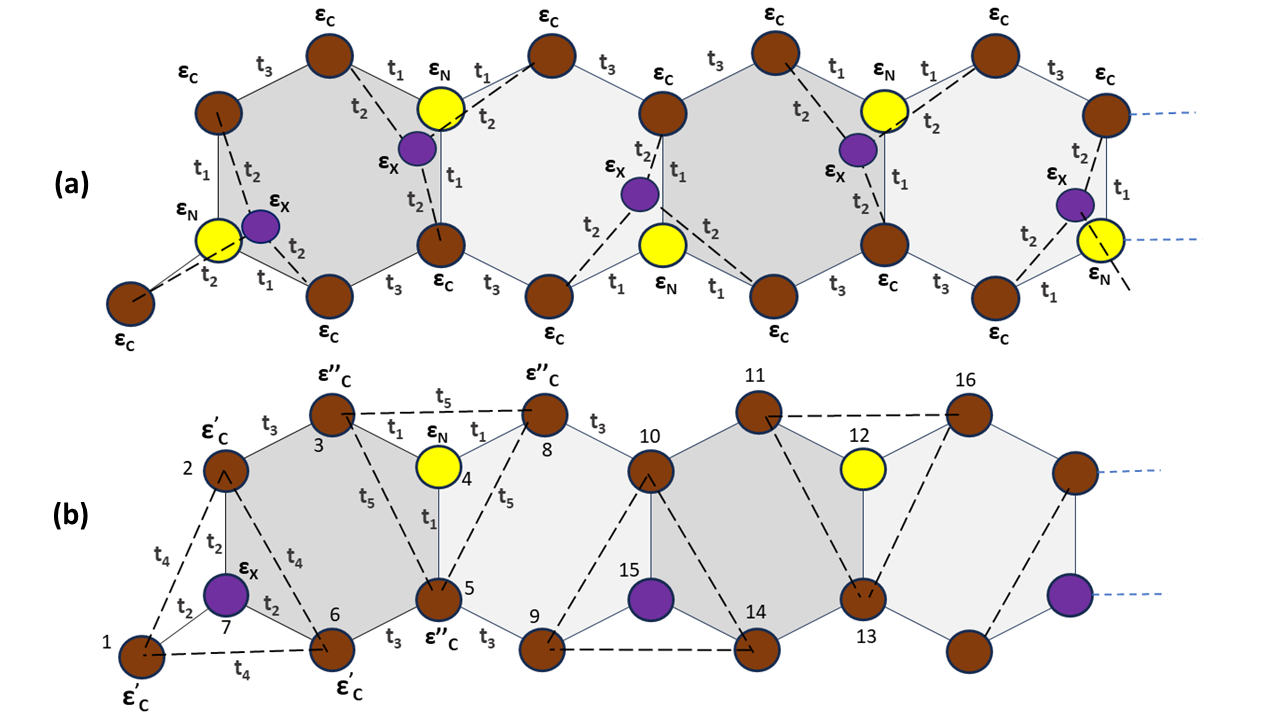}
\caption{\textbf{Nanoribbon and its decimated geometry with corresponding on-site potential energies and hopping parameters.} A portion of an infinitely long ribbon geometry (figure-a) and the decimated \textit{effective} picture of the ribbon (figure-b). Brown, yellow and violet colored balls represent in-plane carbon, buckled nitrogen and the adatoms respectively. The dotted lines in fig-(a) stand for the out-of-plane bonds of these DB C$_3$NX systems.}  
\label{deci}
\end{figure*}
The decimation scheme is pictorially represented in Fig.~\ref{deci}. The lower panel denotes the effectively transformed structure with the corresponding parameters.
If we choose $\epsilon_C$, $\epsilon_N$, and $\epsilon_{Si}$ respectively as the
on-site terms for the carbon, nitrogen and silicon atoms and the hopping terms
$t_{C-N}=t_1$, $t_{C-C}=t_3$ and $t_{C-Si}=t_2$, then in terms of these original
parameters, the effective parameters can be described as follows,
\begin{eqnarray}
t_4 &=& \frac{t_1^2}{(E-\epsilon_N)} \nonumber\\
t_5 &=& \frac{t_2^2}{(E-\epsilon_{Si})} \nonumber\\
\epsilon_C^{'} &=& \epsilon_C + t_4 \nonumber\\
\epsilon_C^{''} &=& \epsilon_C + t_5
\label{decimate}
\end{eqnarray}
These renormalized parameters constitute the \textit{effective version} of the original ribbon.
The new parameters are indicated clearly in the diagram (Fig.~\ref{deci}).
It is needless to mention that the scheme including
Eq.~\eqref{decimate} is not at all sensitive on the numerical values of the kinetic and potential parameters of the Hamiltonian.
This transformed structure remains intact in respect of its periodicity.
\subsection{Spectral information}
\begin{figure}
\centering
\includegraphics[clip,width=7.5 cm,angle=0]{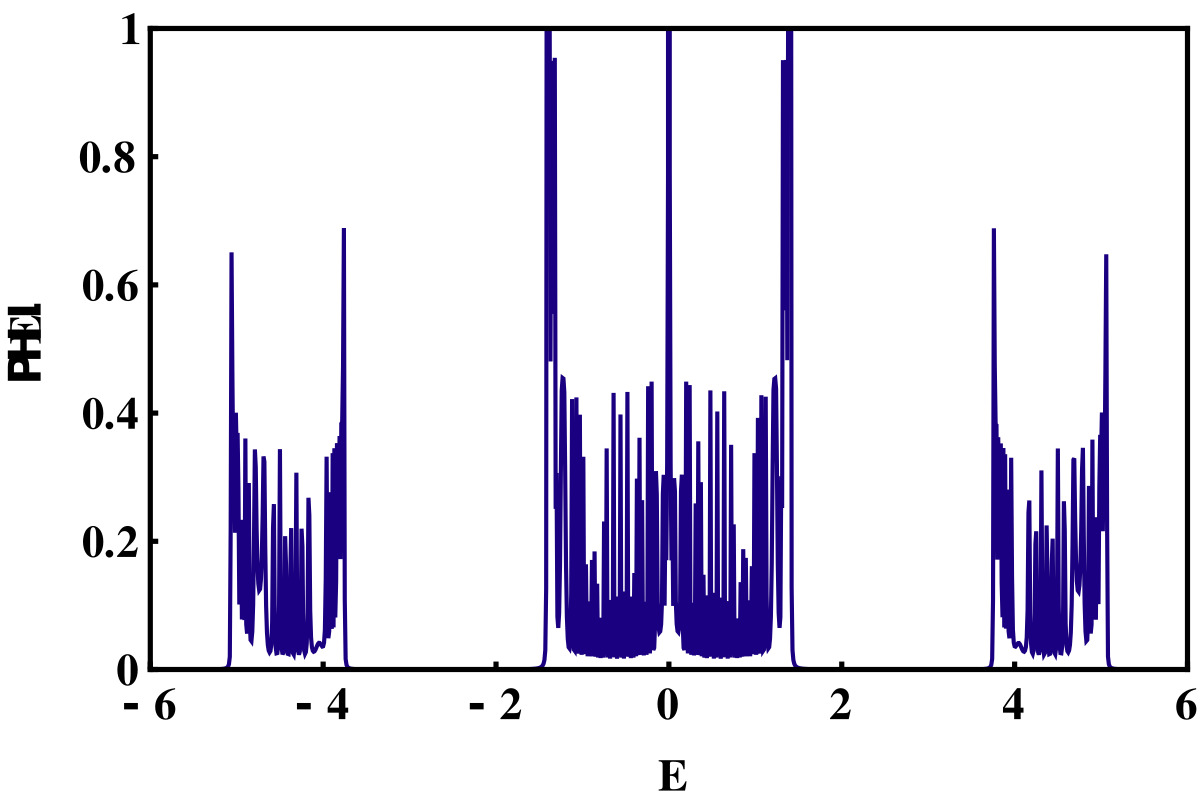}
\caption{\textbf{The variation of density of states with the energy of the incoming projectile.} Here on-site potential of every kind of sites are uniformly set as zero
and the hopping parameters are $t_{C-N}=1.5$, $t_{C-C}=1.0$ and $t_{C-Si}=2.0$.}  
\label{dos}
\end{figure}

Before going to the band dispersion and other associated ideas, we proceed to
acquire an overall idea of the eigenspectrum that describes the entire landscape of allowed eigenmodes as a
function of the energy of the injected projectile. The above effective parameters (Eq.~\eqref{decimate})
will help us to compute the electronic density of states with the help of the standard Green's function technique.
One can easily write down the Hamiltonian matrix
in real space language for a large enough system size to calculate the Green's function.
It is well-known that
the numerical evaluation of average DOS is governed by the following relation, viz.,
\begin{equation}
\rho = -\left( \frac{1}{N \pi} \right) Im[Tr G(E)]
\label{density}
\end{equation}
Where $G(E)=(E-H+ i \delta)^{-1}$ is the usual Green's function and $\delta$ is the negligibly small imaginary damping part added to the energy for the numerical calculation of DOS.
$N$ denotes the system size and `$Tr$' is the trace of the Green's function.
Throughout the numerical analysis, we have set the on-site energy of every kind of
site as, $\epsilon = 0$. This evaluation will give us knowledge about the entire eigenspectrum, the allowed bulk states offered by the ribbon structure and their nature too. 
Nevertheless, as previously outlined, our emphasis will be directed towards analyzing the eigenspectrum of these nanoribbons, with particular attention given to the specific electron-doped regions, owing to the incorporation of N atoms.  
The intricate variation of DOS with respect to the energy is presented in the Fig.~\ref{dos}.
Here we have taken the system size $N=720$.
The portrait of DOS with energy, as shown in the Fig.~\ref{dos}, reveals the formation of \textit{absolutely continuous} subbands densely populated by extended type of eigenfunctions, as expected. There are three such continuum zones present in the DOS plot. We are particularly interested in regions far from charge neutrality that reside within the conduction band of the eigenspectrum obtained from our model. The DOS in the energy region mentioned above serves as the actual DOS of the nanoribbon near the Fermi level because of the artifact of considering uniform on-site energy. The
centro-symmetric pattern is constituted by the transparent modes for which the flow of the hopping parameter shows an oscillatory behavior.
In other words, extensive numerical
search reveals that
for any of the energy belonging to the continuum zone, the non-vanishing overlap integral
demands the non-zero overlap of the wave functions between the consecutive nodes for such modes.
This
confirms the resonant character of the corresponding states. This existence of such Bloch state with
high transmittivity is particularly appreciable due to the translational ordering
 present in the underlying system.
Before ending this demonstration, we should highlight that a central spiky state is
observed at $E=0$. This may associate non-dispersive nature; but since the state lies
in between the continuum zone of diffusive modes, it will loose its in-built localization character. The
reason is self-explanatory.
\subsection{Transport characteristics}
To validate our findings we will now do parametric calculation to explain the transmission behavior of the ribbon-shaped quasi-one dimensional structure with the aid of the usual method of multi-channel transport. For this estimation, we need to clamp the system within a pair of semi-infinite periodic leads. To obtain the transmittance, here we now
embrace the standard Green's function approach.
\begin{figure}
\centering
\includegraphics[clip,width=7.5cm,angle=0]{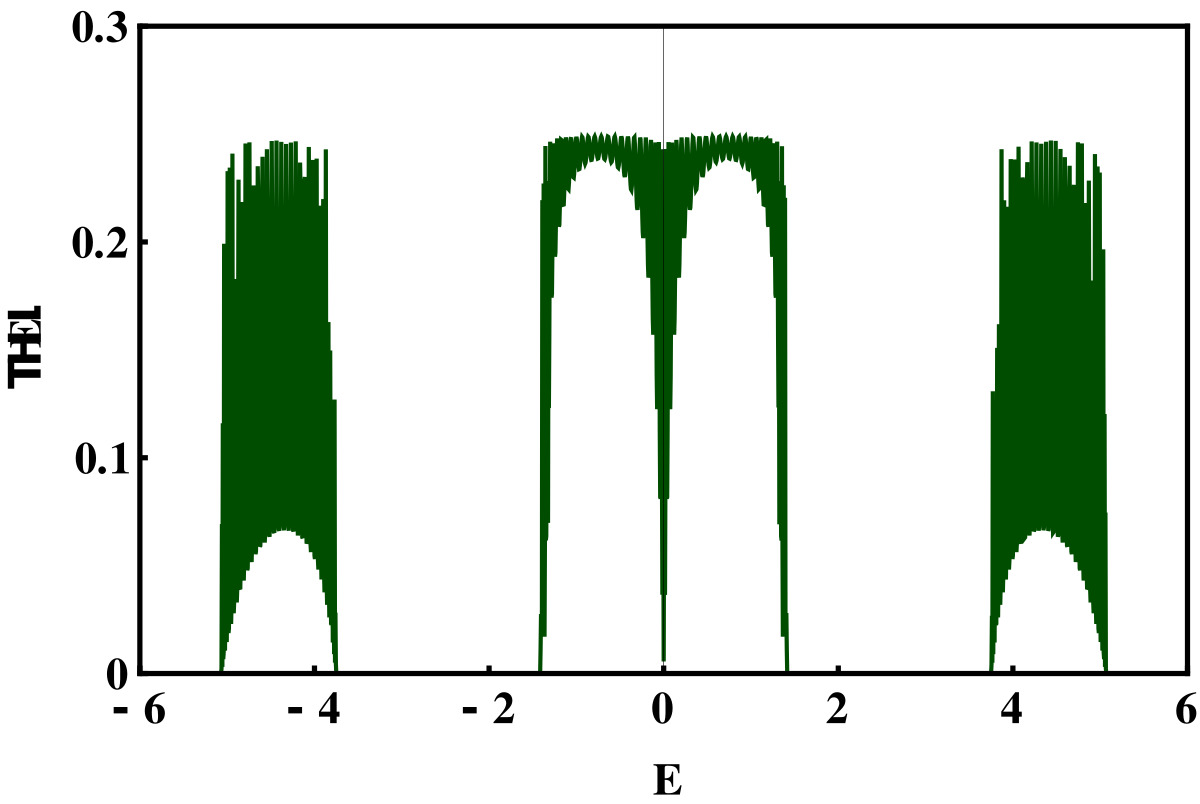}
\caption{\textbf{The variation of transmission
probability with respect to the energy of the
incoming excitation.} Here the on-site potential of every kind of sites are uniformly set as zero
and the hopping parameters are $t_{C-N}=1.5$, $t_{C-C}=1.0$ and $t_{C-Si}=2.0$.}  
\label{trans}
\end{figure}
The graphical variation of the transmission probability is depicted in the Fig.~\ref{trans}. From this
pattern it is clear that there are three identical resonant zones which are
essentially compatible with the DOS plot. For the modes residing over the
continuum zones, it shows resonating transport and this confirms the diffusive
character. The total number of sites taken for this numerical evaluation is $N=160$.
The parameters of the Hamiltonian are necessarily set as identical with that considered in the
calculation of DOS.
The on-site energy of the atomic sites of the semi-infinite leads is set as zero and the hopping
is put as $t_L=3$ to obtain the full band picture. Also, the system to lead coupling
is taken as identical with $t_L$ to minimize the scattering probability at the connectors. For the central spiky mode, there is a drastic drop of transmittance
due to the relatively low localization length of the state.
The ballistic transmission of the states of the other two zones (left and right) is evident
from the diagram. The conducting nature of those modes automatically validates
the expected diffusive flavor
of the Bloch states. Similar to the previous section, the Fermi level of the original system lies at the conduction band, and the corresponding transmittance of the system will only be observed in practice.   
\subsection{Topological phase transition}
\begin{figure*}
    \centering
    \includegraphics[scale=0.46]{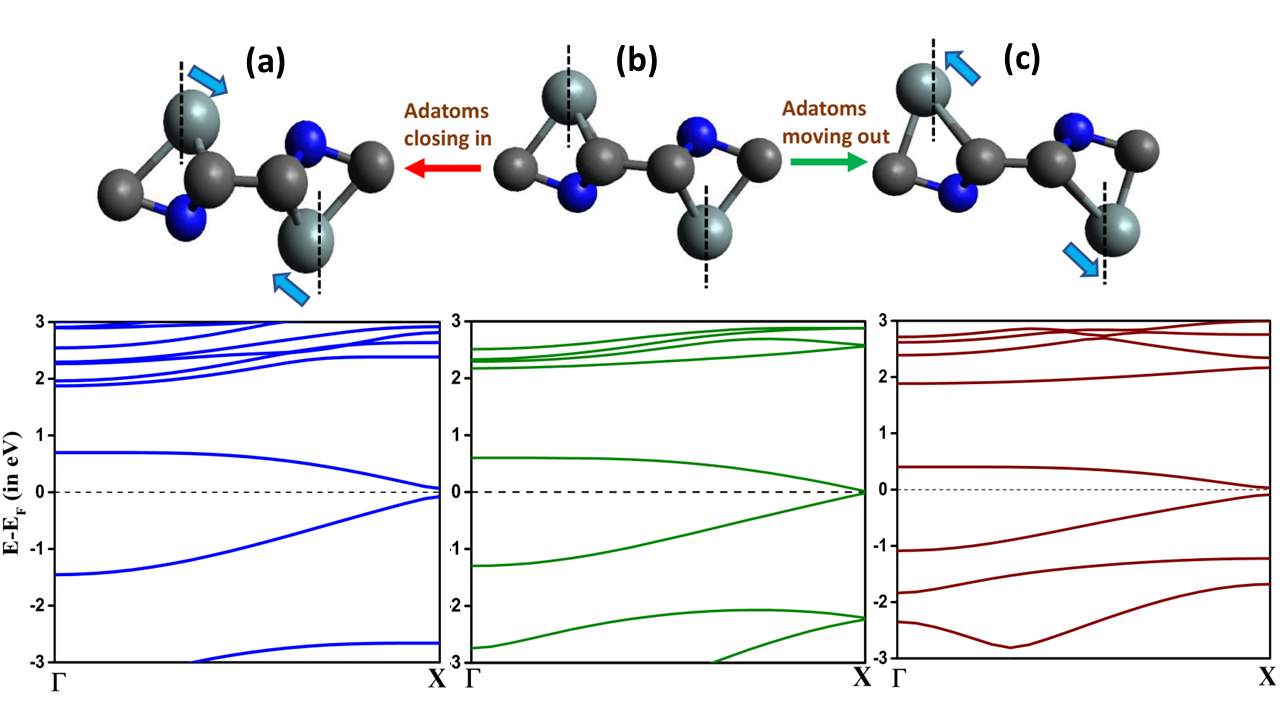}
    \caption{\textbf{Nanoribbon form of dumbbell C$_3$NSi with different types of twists and their corresponding electronic nature}. The optimization of these edge-passivated ribbons and the subsequent band structures are found using DFT. The dotted lines in all three figures show the relaxed position of the adatoms. Here figures (a) and (c) show the formation and the band structure of the ribbon under in-phase inward (where the adatoms move close to each other) and outward (move far apart) strain whereas figure-(b) depicts the structure and the electronic nature of the pristine nanoribbon. The critical phase is gapless while other twisted phases are gaped with distinct topological properties.}
    \label{fig:ribbonband}
\end{figure*}

As we discussed earlier, the low-energy physics of these quasi-one-dimensional DB C$_3$NX systems can be effectively represented by considering an effective network comprised of ``X" atoms [Fig.~\ref{fig:ribbonband}]. This effective system bears a resemblance to a one-dimensional chain of atoms with two sub-lattices per unit cell. However, a careful examination unveils a uniformity in the bond lengths of ``X" atoms, both within intra- and inter-unit cell connections. This additional symmetry implies that the original system can be essentially mapped onto a uniform chain of atoms, where the lattice periodicity constitutes half of the initial lattice parameter. In other words, the uniform one-dimensional lattice geometry reveals that the electronic spectra will invariably exhibit a gap-closed phase unbiased by the specific type of ``X" atoms as shown in Fig.~\ref{fig:ribbonband}(b). The above discussion can be excellently supported by the DOS spectra and transmission probability of the C$_{3}$NX ribbon as discussed in the previous section, which is identical to that of the one-dimensional chain of atoms. 
\begin{figure}
    \centering
    \includegraphics[scale=0.28]{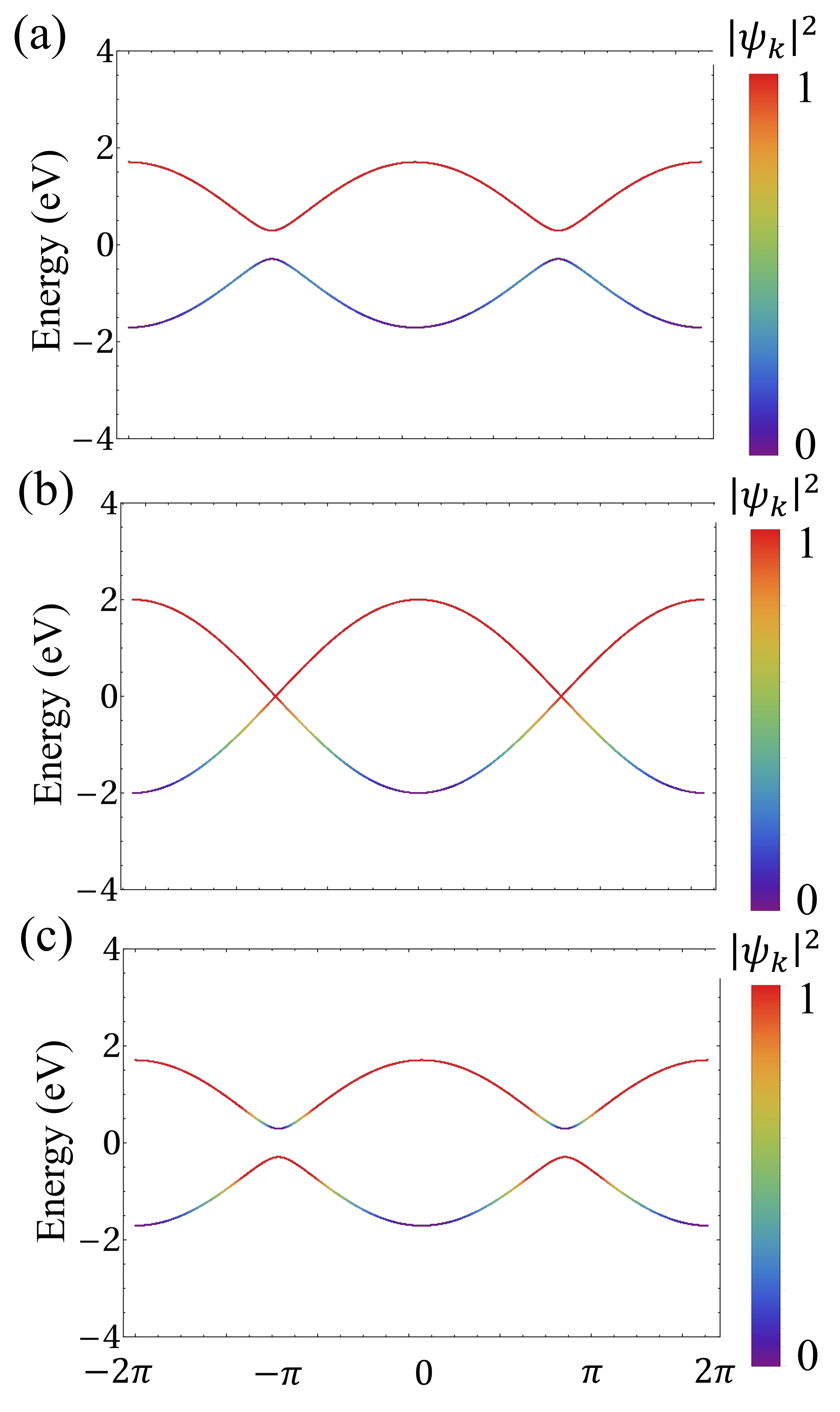}
    \caption{\textbf{Illustration of topological characteristics in C$_3$NX dumbbell configurations, governed by the Su-Schrieffer-Heeger model.} The 1D band dispersion of a two-band Hamiltonian with X atoms (X = C, Si, Ge) and $\abs{\psi_{k}}^2$ contribution is displayed. (a) Stronger inter-unit cell X atom hopping results in a trivial insulator band gap ($w = 1/\sqrt{2}$, $v = 1$). (b) Equal contributions lead to a semi-metallic band dispersion ($v = w = 1$). (c) Stronger intra-unit cell hopping induces another gap-opened phase ($w=1$, $v = 1/\sqrt{2}$), highlighting orbital exchange during the topological phase transition. All these evaluations are done using the realistic tight-binding method.}.
    \label{fig:ssh}
\end{figure}
\begin{figure*}
    \centering
    \includegraphics[scale=0.46]{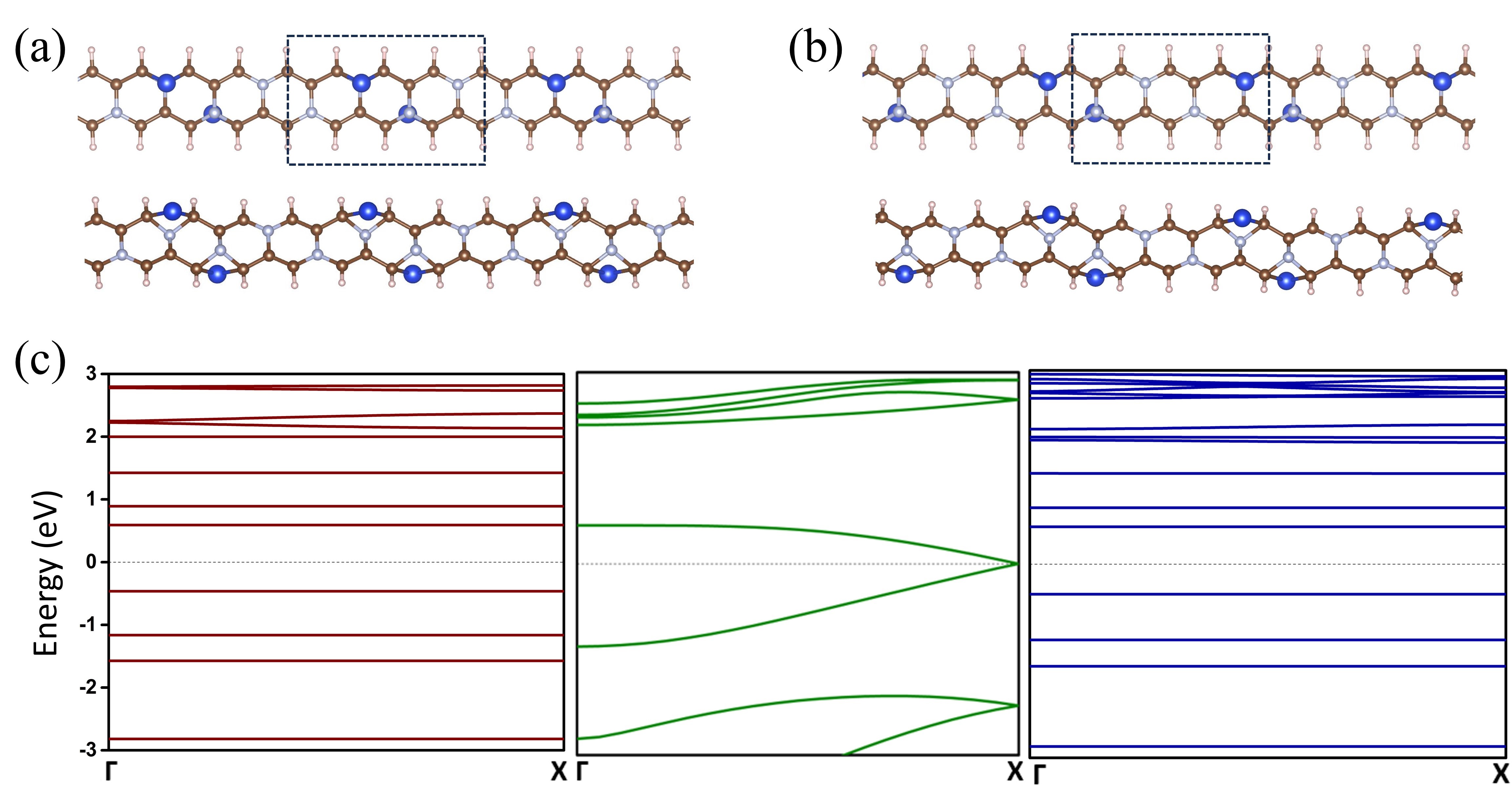}
    \caption{\textbf{Nanoribbon form of dumbbell C$_3$NSi and their fully dimerized limits with the subsequent electronic nature.} The top (upper panel) and side (lower panel) views of DB C$_3$NX nanoribbons with fully relaxed geometries, illustrating (a) the absence of X atoms near the edge of the unit cell and (b) the absence of X atoms at the center of the unit cell. Additionally, (c) presents the DFT band structure: the left panel corresponds to the ribbon in subfigure (a), the middle panel shows the original ribbon without removed X atoms, and the right panel corresponds to the ribbon in subfigure (b). The left and right panels exhibit gapped phases with distinct topological signatures.}
    \label{fig:sshre1}
\end{figure*}
\begin{figure*}
    \centering
    \includegraphics[scale=0.48]{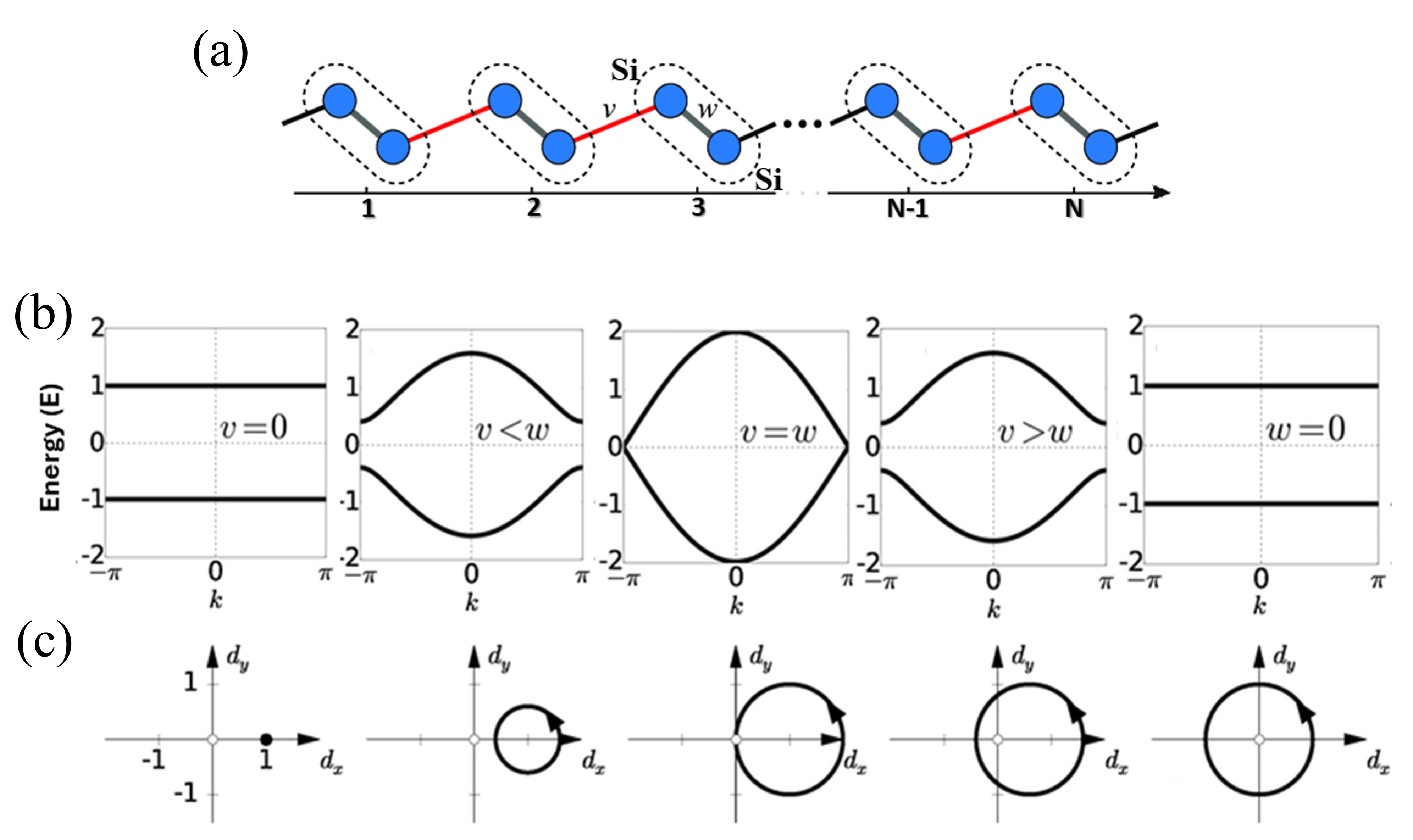}
    \caption{\textbf{Schematic representation of the low-energy DB C$_3$NSi as a quasi-1D network of `Si' with staggered inter- and intra-unit cell hopping (w and v) and the evolution of band structure with underlying winding number}. Figure-(a) shows the schematic of the SSH chain made of the X atoms (X =Si, here), (b) shows the band structure of this SSH chain depending on the different values of the intra-cell (w) and inter-cell (v) hopping parameter and (c) displays the winding number for different conditions reveals that $v<w$ is always topologically trivial whereas $v>w$ are topologically nontrivial.}
    \label{fig:sshre2}
\end{figure*}
Motivated by the structural intricacies of the original lattice geometry and corresponding symmetry, we try to induce a band gap in the system. In particular, we have realized that the gap-closed phase as mentioned above is similar to the critical phase of the SSH model \cite{heeger1988solitons,li2014topological,maffei2018topological,hasan2010colloquium,parto2018edge}. The experimental quantum simulation of this model and realization of localized topological soliton state through quenched dynamics,
phase-sensitive injection, and adiabatic preparation is already reported \cite{meier2016observation}. Therefore, we can essentially break the symmetry of the intra- and inter-unit cell bond lengths to essentially form a two level system. For example, in the case of C$_3$NSi, the Si atoms undergo a shift of 1.91 \AA{} and 2.29 \AA{} compared to the unperturbed case, for tensile and compressive type strain, respectively, as illustrated in Fig.~\ref{fig:ribbonband}. As we have discussed earlier, only p$_z$ orbitals of the X atoms contribute in forming the states near the Fermi level. Therefore, we have mapped this quasi-1D lattice problem to an effective low-energy model composed exclusively of X atoms, which accurately reproduces the band structure near the Fermi level.
\begin{figure*}
    \centering
    \includegraphics[scale=0.46]{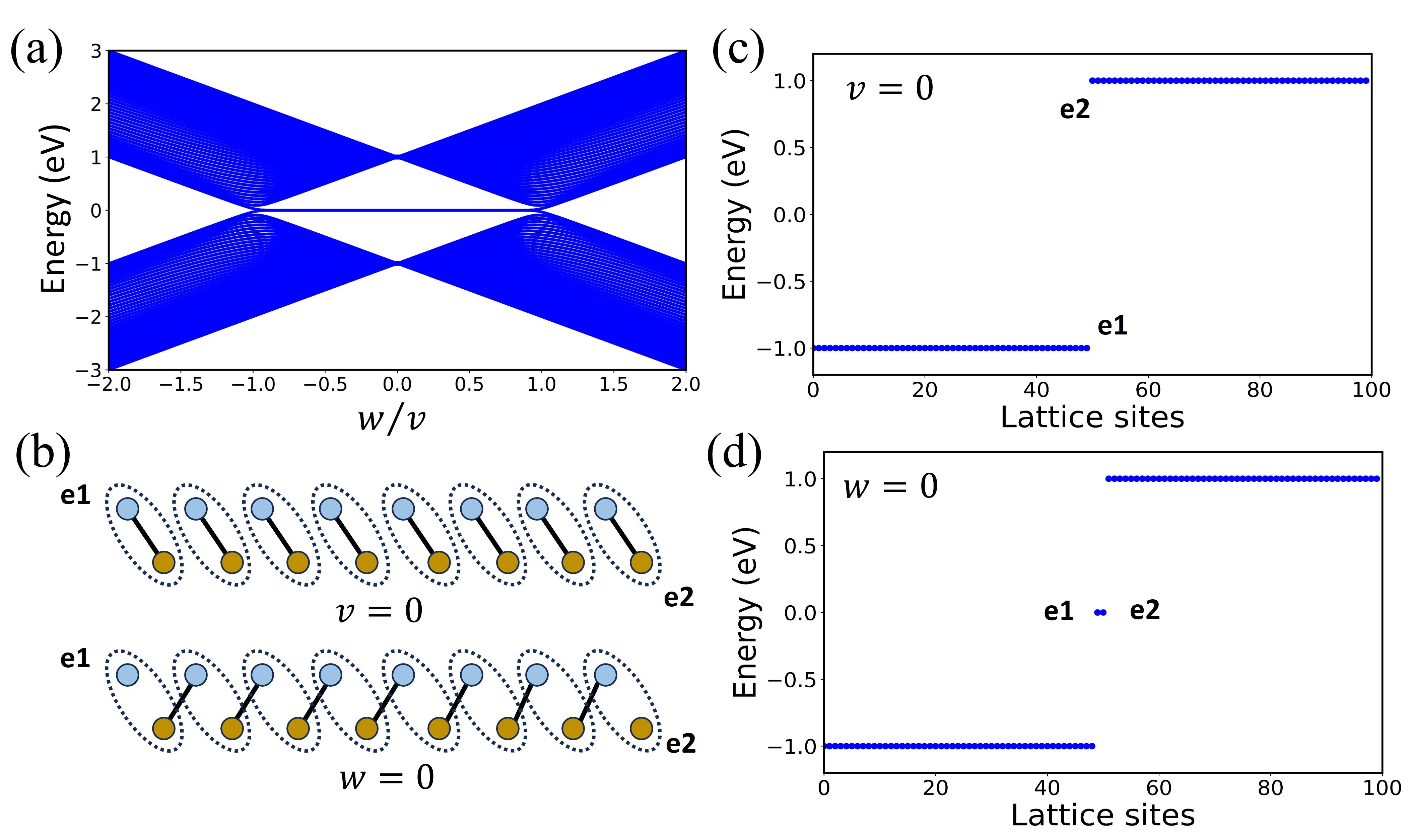}
    \caption{\textbf{Appearance of zero-energy states with its underlying energy spectrum}. Figure-(a) shows the zero energy edge state of the nanoribbon (manifesting the SSH chain) under the open boundary condition for v$>$w. Figure-(b) depicts the schematic representation of the SSH chain for the condition $v=0$ and $w=0$ and (c) shows the absence of the zero energy edge state for the $v=0$. Figure-(d) displays the presence of zero energy edge states for the $w=0$ case. The plots are obtained for the X atoms with 100 lattice sites.}
    \label{fig:sshre3}
\end{figure*}
It is needless to mention that in this case the system behaves as a two-level system with lattice periodicity $a$ characterized by the Hamiltonian
\begin{equation}
    H_{k} = \begin{pmatrix}
0 & w + v \: e^{i k a} \\
w + v \: e^{-i k a} & 0
\end{pmatrix}
\label{eq:sshham}
\end{equation}
In the above eq.~\ref{eq:sshham}, the hopping parameters $w$ and $v$ represent the intra- and inter-unit cell hopping. The above Hamiltonian gives rise to the bulk dispersion relation in terms of energy eigenvalues.
\begin{equation}
    E = \pm \sqrt{v^2 + w^2 + 2 v w \: cos (k a)}
    \label{eq:expresson_ssh}
\end{equation}
Similarly, the eigenstates, or eigenvectors, exhibit the following form:
\begin{eqnarray}
    \ket{\pm k} = \frac{1}{\sqrt{2}}\begin{pmatrix}
\pm e^{-i \lambda(k)} \nonumber \\
1
\end{pmatrix}, \\
\lambda (k) = tan^{-1} \left(\frac{v \: sin (k)}{w + v \: cos (k)}\right).
\end{eqnarray}
In the critical phase, two hopping parameters $w$ and $v$ become equal, and we obtain a gapless phase similar to the zone-folded uniform chain of atoms. On the other hand, different hopping entropy ($w \neq v$) opens a gap in the system that belongs to different topological phases. For example, Fig.~\ref{fig:ssh} illustrates that both $v < w$ and $v > w$ open a band gap; however, the orbital exchange in the latter case establishes the nontrivial topological phase with winding number $\nu = 1$. In Fig.~\ref{fig:ssh}, we have considered two representative hopping parameters $1$ and $1/\sqrt{2}$ to explain the distinct topological nature of two gap opened phase $v < w$ and $v > w$. It is not to mention the physics of a particular phase remain unaltered as long as the corresponding hopping relation is protected. A very recent report on the dumbbell graphene proves the similar adjusting of the atomic positions of adatoms which can tune the topological property of the system \cite{song2024hydrogenation}.\\
Motivated by this observation, we have applied a shearing strain to the system, inducing a tilt in the dumbbells formed by the X atoms within the nanoribbon structure. We observed that two different types of shearing strain: (i) causing two X atoms of a unit cell to come closer to each other, and (ii) causing two X atoms of a unit cell to move farther apart from each other, result in gap opened phases but of differing topological natures. Under the application of strain, these geometries are the ground state and the corresponding band structures are obtained using DFT. However, addressing this issue, we have further explored the possibility of obtaining distinct topological phases in the DB C$_3$NX nanoribbon without any strain engineering. We note that in the DB C$_3$NX nanoribbon, the dumbbells are formed on the top of each N atom. It is extremely interesting that if the concentration of the dumbbells are reduced to its half in two distinct ways given in Fig.~\ref{fig:sshre1}(a) and (b) we obtain two district topological phases in the normal ground state, i.e., without any external perturbation. The emergence of these distinct non-trivial phases can be properly explained in terms of the SSH model. The above statement is underpinned by the fact that only the X-atoms predominantly contribute near the Fermi level. Therefore, in a low-energy depiction this ribbon is identical to a SSH chain consists of two X atoms per unit cell placed on the top of each N atom. The band structure of this unperturbed nanoribbon is semimetallic with a zero band gap, as shown in Fig.~\ref{fig:sshre1}(c) (middle panel). Now, we have reduced the concentration of the X atoms forming the dumbbell in the ribbon geometry. The above strategy can be attained by constructed a $2\times 1$ supercell, which consists of four X atoms on the top of four N atoms. First, we remove two of the four X atoms located at the edge of the unit cell, ensuring that the X atoms at the center of the unit cell remain intact. Therefore, we realize that the distance (d) between the intracell X atoms is very less compared to that of the intercell X atoms [Fig. \ref{fig:sshre1}(a)]. As a consequence, the hopping parameter between the intracell X atoms (w $\propto d^{-2}$) is stronger compared to that of the intercell one (v) of the SSH chain [Fig.~\ref{fig:sshre2}(a)]. On the other hand, removing two central X atoms inside the unit cell leads to the structure given in Fig.~\ref{fig:sshre1}(b), where the inter-unit cell bond length (hopping) is less (stronger) than that of the intra-unit cell one. We have computed the DFT band structure of these two systems given in Fig.~\ref{fig:sshre1}(c) (left panel) and (right panel). It is evident that the band structure near the Fermi energy will be identical to the band dispersion of the SSH chain given in eq.~\ref{eq:expresson_ssh}. Therefore, there will be always a gap in the band structure unless $v=w$, given in Fig.~\ref{fig:sshre2}(b). Now, we see when the $v$ is negligible compared to w (similar, $v=0$) or w is negligible compared to v (similar, $w=0$) the coefficient of cos(⁡ka) vanishes and we obtain a flat band. However, the calculated winding number illustrated in Fig.~\ref{fig:sshre2}(c) reveal that v=0 condition is topologically trivial whereas the w=0 condition is topologically nontrivial. It is needless to mention that the flat bands near the Flat band, shown in Fig.~\ref{fig:sshre1}(c) can be explained using the above discussion. Furthermore, the Fig.~\ref{fig:sshre1}(c) (left panel) is topologically trivial whereas the Fig.~\ref{fig:sshre1}(c) (right panel) is topologically nontrivial.\\
Moreover, in order to support the above statement, we have performed calculation of the system under open boundary condition. We have observed that the edge states are present in the open boundary condition for $w<v$, including $w=0$ case [Fig.~\ref{fig:sshre3}(a)]. Two different types of edges under open boundary condition are illustrated in Fig.~\ref{fig:sshre3}(b). The absence and presence of zero energy modes for $v=0$ and $w=0$, respectively, as given in Fig.~\ref{fig:sshre3}(c) and Fig.~\ref{fig:sshre3}(d) critically establishes the topologically trivial and nontrivial phase for $v=0$ and $w=0$, respectively. Therefore, the above discussion reveals that the nanoribbon shown in Fig.~\ref{fig:sshre1}(a) is topologically trivial, whereas the nanoribbon shown in Fig.~\ref{fig:sshre1}(b) is topologically nontrivial in its pristine form.\\
From the above discussion, we note that the SSH chain, characterized by alternating strong and weak bonds, serves as a paradigmatic model for understanding topological phases. This model has been naturally observed in the polyacetylene polymer chain. In this context, the uniform chain of carbon atoms spontaneously adopts one of the possible ground states with alternating bond lengths, which can be either topologically trivial or topologically non-trivial. In the case of the DB C$_3$NX system, we have explored a compelling depiction that mimics similar SSH physics in the low-energy spectrum. This has been critically established using the projected density of states near the Fermi level and the signature of band inversion subjected to systematic lattice distortion. The effective low-energy lattice of the DB C$_3$NX nanoribbons consists of two X atoms and is essentially underpinned by the same arguments introduced for the paradigmatic SSH chain. Furthermore, the low-energy version of the original lattice can be achieved under the influence of strain engineering on the original nanoribbon, as introduced here.

\section{Conclusion}
In summary, we have found compelling non-trivial topological results in the recently proposed stable dumbbell-like sp$^3$ hybridized carbon nitride systems namely DB C$_3$NX (X= C, Si, Ge). The emergence of the Dirac cone around the Fermi energy level is solely due to the presence of adatoms which further led us to construct an effective tight-binding model Hamiltonian that explored low-energy physics around the Fermi level. Comparable spin-orbit interaction strength to Xenes, these systems display a finite gap opening for DB C$_3$NSi (and Ge) - showing the signs of a quantum spin hall (QSH) insulator. Further presence of a considerable amount of buckling permits us to apply an external transverse electric field that breaks the inversion symmetry. Beyond a critical electric field, these materials depict a sharp topological phase transition. Berry curvature plot on the bands near the Fermi energy level properly explains the closing of this band gap with a clear band inversion around the critical field value. We have made this topological identity more profound by calculating the topological index $\mathbb{Z}_2$, which is 1 in this case, using Wannier charge centers (WCC). Next, we have investigated the overall spectral contribution and transport attributes of the DB C$_3$NX nanoribbons using the real space decimation technique and the Green function method. Explicit traces of topological phases are found using the famous Su-Schrieffer-Heeger (SSH) model where the in-phase outward motion of the adatoms leads to non-trivial consequences.
\section*{Declaration of competing interest}
The authors declare that they have no known competing financial interests or personal relationships that could have appeared to influence the work reported in this paper.
\section*{CRediT authorship contribution statement}
\textbf{DM}: Conceptualisation, Computations, Formal Analysis, Writing original draft. \textbf{AB}: Conceptualisation, Computations, Formal analysis, Writing original draft. \textbf{AN}: Computations, Formal Analysis, Writing original draft. \textbf{DJ}: Conceptualisation, Supervision, Finalizing manuscript.
\section*{Acknowledgements}
D.M. sincerely acknowledges the Council of Scientific and Industrial Research (CSIR) India for providing fellowship and his co-authors for constructive discussions. A.B. sincerely thanks the financial support from the Indian Institute of Science IoE postdoctoral fellowship. 
\bibliographystyle{elsarticle-num}
\bibliography{ref.bib}
\end{document}


\begin{frontmatter}
\title{Quantum Spin Hall Effect and Su-Schrieffer-Heeger Model Implementation in Novel C$_3$N-based Dumbbell Morphologies}
\author[1]{Deep Mondal\corref{cor1}}
\author[2]{Arka Bandyopadhyay\corref{cor1}}
\author[3]{Atanu Nandy}
\author[1]{Debnarayan Jana\corref{ac}}
\ead{djphy@caluniv.ac.in}
\cortext[cor1]{These authors contributed equally to this work}
\cortext[ac]{Corresponding author}
\address[1]{Department of Physics, University of Calcutta, 92 A. P. C. Road, Kolkata-700009, India}
\address[2]{Solid State and Structural Chemistry Unit, Indian Institute of Science, Bangalore 560012, India}
\address[3]{Department of Physics, Acharya Prafulla Chandra College, New Barrackpore, Kolkata 700131, India}
\end{frontmatter}
\begin{figure*}[ht]
    \centering
    \includegraphics[width=16.5 cm]{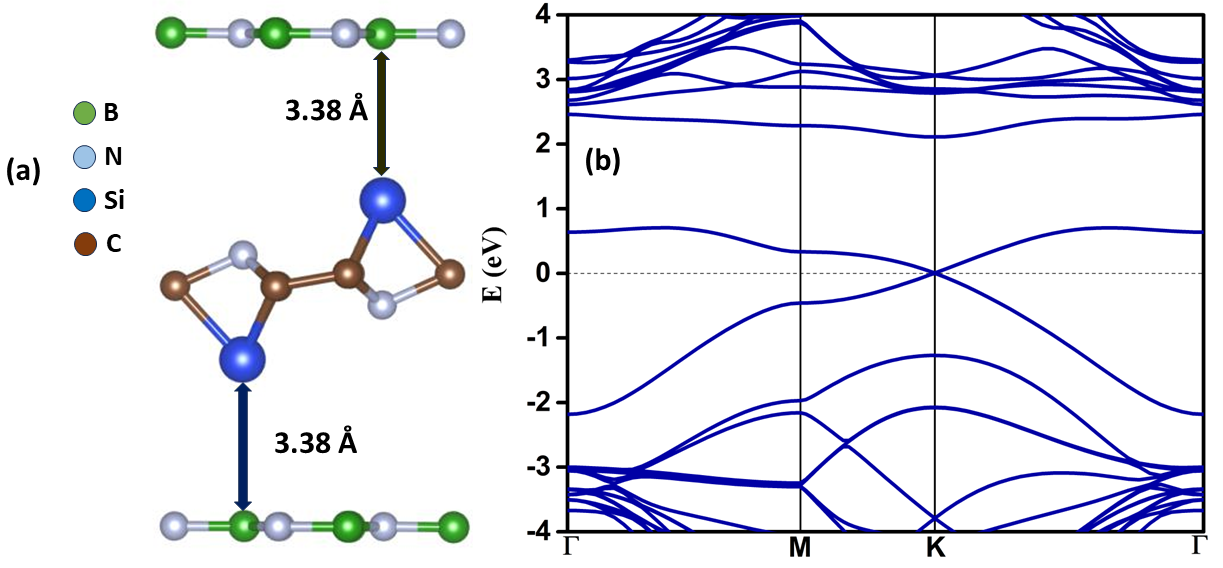}
    \caption{\textcolor{blue}{\textbf{Relaxed unit cell of hBN/C$_3$NSi/hBN heterostructure and its electronic nature}. Here figure (a) shows the optimized construct and (b) shows the electronic structure. The C (brown), N (fade blue), Si (blue) and B (green) are the carbon, nitrogen, silicon and boron atoms respectively.}} 
    \label{hetero}
\end{figure*}
\begin{figure*}[ht]
    \centering
    \includegraphics[width=14.5 cm]{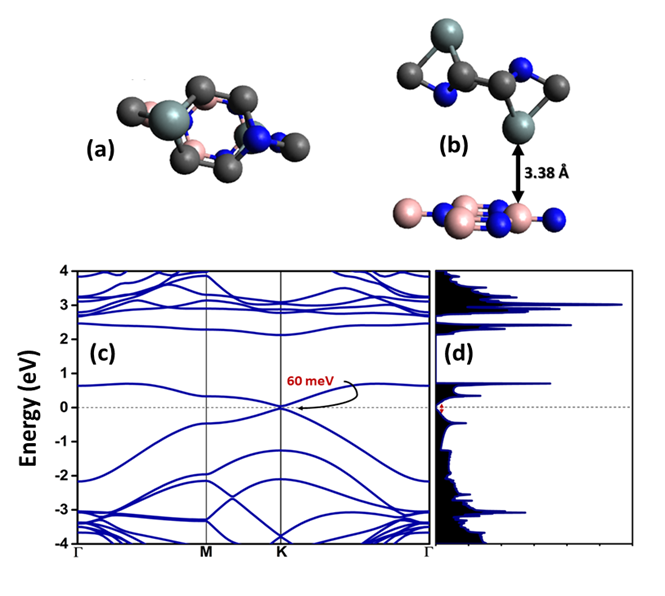}
    \caption{\textbf{Relaxed unit cell of C$_3$NSi/hBN heterostructure and its electronic nature}. Here figure (a) and (b) shows the optimized construct from top and side views and (b) shows the electronic structure and the density-of-states.} 
    \label{hetero1}
\end{figure*}
\begin{figure*}[ht]
    \centering
    \includegraphics[width=16.5 cm]{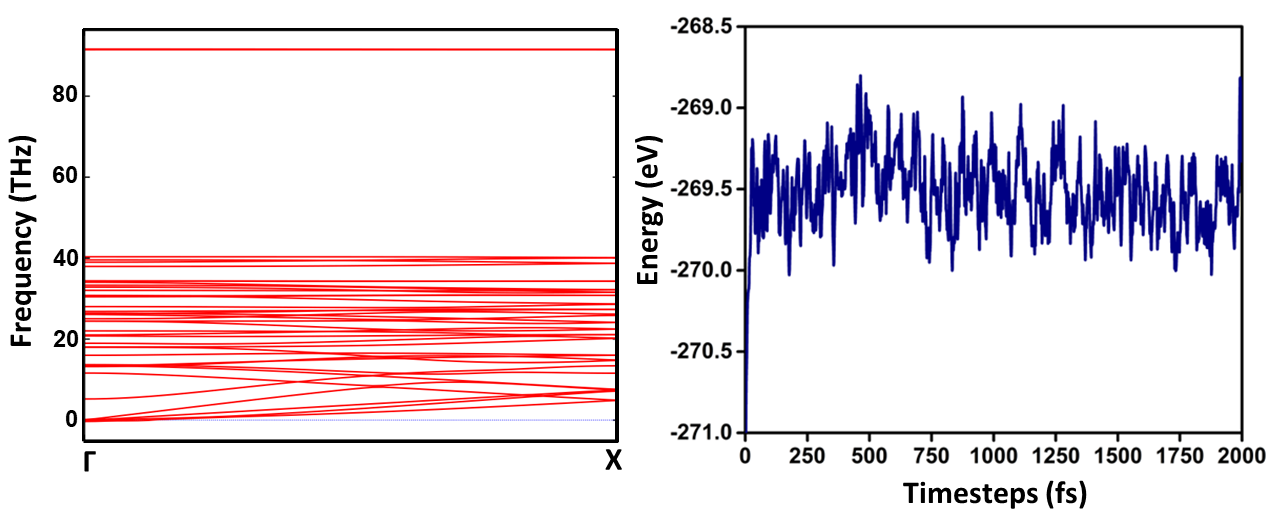}
    \caption{\textbf{Phonon dispersion and molecular dynamics simulation (300 K) curve of DB C$_3$NSi nanoribbon.}} 
    \label{MD_ribbon}
\end{figure*}